\documentclass{iopart}
\usepackage{graphicx}
\usepackage{amssymb}

\begin{document}

\title{Skyrmion Dynamics and Topological Sorting on Periodic Obstacle Arrays}
\author{N.P. Vizarim$^{1,2}$, C. Reichhardt$^{1}$, C.J.O. Reichhardt$^{1}$, and P.A. Venegas$^{3}$}
\address{$^{1}$Theoretical Division and Center for Nonlinear Studies,
Los Alamos National Laboratory, Los Alamos, New Mexico 87545, USA}
\address{$^{2}$ POSMAT - Programa de P{\' o}s-Gradua{\c c}{\~ a}o em Ci{\^ e}ncia e Tecnologia de Materiais, Faculdade de Ci\^{e}ncias, Universidade Estadual Paulista - UNESP, Bauru, SP, CP 473, 17033-360, Brazil
  }
\address{$^{3}$ Departamento de F\'{i}sica, Faculdade de Ci\^{e}ncias, Universidade Estadual Paulista - UNESP, Bauru, SP, CP 473, 17033-360, Brazil
}
\ead{cjrx@lanl.gov}

\begin{abstract}
  We examine skyrmions under a dc drive interacting with a square array of obstacles for varied obstacle size and damping. When the drive is applied in a fixed direction, we find that the skyrmions are initially guided in the drive direction but also move transverse to the drive due to the Magnus force.   The skyrmion Hall angle, which indicates the difference between the skyrmion direction of motion and the drive direction, increases with drive in a series of quantized steps as a result of the locking of the skyrmion motion to specific symmetry directions of the obstacle array.  On these steps, the skyrmions collide with an integer number of obstacles to create a periodic motion.  The transitions between the different locking steps are associated with jumps or dips in the velocity-force curves.  In some regimes, the skyrmion Hall angle is actually higher than the intrinsic skyrmion Hall angle that would appear in the absence of obstacles.  In the limit of zero damping, the skyrmion Hall angle is 90$^\circ$, and we find that it decreases as the damping increases.  For multiple interacting skyrmion species in the collective regime, we find jammed behavior at low drives where the different skyrmion species are strongly coupled and move in the same direction.  As the drive increases, the species decouple and each can lock to a different symmetry direction of the obstacle lattice, making it possible to perform topological sorting in analogy to the particle sorting methods used to fractionate different species of colloidal particles moving over two-dimensional obstacle arrays.
\end{abstract}

\maketitle

\vskip 2pc

\section{Introduction}
For a particle moving over a two dimensional periodic array of scattering sites, the type of motion
that occurs depends on the direction of drive relative to the symmetry directions of the array.  For certain drive angles,
the particle can follow a straight trajectory without encountering any obstacles,
while for other angles, the particle
collides unavoidably with the obstacles.
Such systems can exhibit
a directional locking effect in which particles preferentially
channel along certain directions. 
In a square lattice of obstacles,
these angles include
$\phi=0^\circ$, $45^\circ$ and $90^\circ$,
and they are more generally described by the relation
$\phi = \arctan(n/m)$ where $n$ and $m$ are integers
\cite{Reichhardt99,Wiersig01,Korda02,Reichhardt12}. 
Dynamical directional locking effects have been studied for vortices interacting with 
square and triangular pinning arrays
\cite{Reichhardt99,Silhanek03,Reichhardt08a},
electrons moving through antidot arrays \cite{Wiersig01,Khoury08} 
and colloidal assemblies moving on two dimensional periodic substrates
\cite{Korda02,Reichhardt12,Gopinathan04,Reichhardt04,MacDonald03,Lacasta05,Roichman07a,Herrmann09,Li19}.
In these systems, 
as the angle of the drive is varied with respect to the symmetry direction of the
substrate, the velocity vector or velocity-force curves
show a series of steps corresponding to
drive angle intervals over which the direction of the motion of the
particles remains locked to the substrate instead of following
the drive direction.
The steps are similar to the phase locking 
phenomenon studied in systems of particles moving in a fixed direction over 
a periodic substrate under combined ac and dc driving.
When phase locking occurs,
the frequency of the oscillations generated by the motion of the particle
over the periodic substrate
locks or comes into resonance with
the ac drive frequency and its higher harmonics
for a fixed range of drive intervals, producing steps
in the velocity-force curves of the type found in Josephson junctions 
(which are known as Shapiro steps) \cite{Shapiro63,Benz90},
incommensurate sliding charge density waves \cite{Coppersmith86},
driven Frenkel-Kontorova systems \cite{Hu07},
vortices in type-II superconductors moving over
a periodic pinning substrate
\cite{Martinoli78,Reichhardt00b,Dobrovolskiy15},
and colloidal particles driven over periodic substrates \cite{Juniper15,Brazda17}. 
In the case of the directional locking,
there is no ac driving; however, two frequencies are still present, where
one is associated with motion in the direction parallel to the drive and the other is
associated with motion in the direction perpendicular to the drive.
Directional locking can also arise in
a system with a quasiperiodic substrate, where there are five
or seven symmetry locking directions \cite{Reichhardt11,Bohlein12a}.
The locking can be harnessed for applications
such as the sorting of different species of particles which
have different sizes, charges, or damping,
where a spatial separation of the species is achieved over time when
one species locks to one angle while the other species locks to
a different angle
\cite{Huang04,Ladavac04,Jonas08,Xiao10a,Tahir11,Risbud14,Wunsch16}. 
Up to this point,
directional locking effects have been studied solely in
overdamped systems where the transitions between directionally locked
states can only occur
when the direction of the drive with respect to the substrate symmetry is
changed.

Skyrmions in chiral magnets are another type of particle-like system
with distinctive properties
\cite{Muhlbauer09,EverschorSitte18,Yu10,Nagaosa13}.
Magnetic skyrmions 
have been found in a wide variety of materials with
skyrmion sizes ranging from a micron down to 10 nm, and
in a number of materials,
the skyrmions are stable at room temperature \cite{Woo16,Soumyanarayanan17}.
Skyrmions are readily set into motion
via the application of an external current
\cite{Nagaosa13,Schulz12,Yu12,Iwasaki13,Lin13a},
and the resulting velocity-force relations show a pinned to sliding
transition that can be observed in
transport experiments
by measuring changes in the topological Hall effect \cite{Schulz12,Liang15} 
or performing direct imaging of the skyrmion motion
\cite{Woo16,Yu12,Montoya18}.
It is also possible to examine
skyrmion dynamics using neutron scattering \cite{Okuyama19},
X-ray diffraction \cite{Zhang18}, 
and changes in the noise fluctuations as a function of drive \cite{Diaz17,Sato19}.

Due to their stability, size scale, and manipulability, skyrmions
are very promising candidates for a variety of
applications including memory, logic devices, and alternative computing architectures
\cite{Fert17,Prychunenko18}.
The capability to
precisely control the direction, traversal distance, and reversibility of
skyrmion motion
could open up new ways to create
such devices,
and there are already a number of proposals
for controlling skyrmion motion using
structured substrates
such as race tracks 
\cite{Fert17,Navau16,Leliaert19}, 
periodic modulations \cite{Reichhardt16a},
or specially designed pinning structures \cite{Ma16,Ma17,Stosic17,Fernandes18,Toscano19}. 
One proposal for controlling skyrmion motion
involves having the skyrmions
interact with a two dimensional periodic substrate of the type that has
already been
realized for colloidal particles and vortices in type-II superconductors,
and there are existing experimental realizations of skyrmions 
interacting with two-dimensional (2D) anti-dot arrays \cite{Saha19}.

A key feature that distinguishes skyrmions
from colloids and superconducting vortices is the 
strong non-dissipative Magnus component in the skyrmion dynamics
caused by topology \cite{EverschorSitte18,Nagaosa13,EverschorSitte14}.
This affects both
how the skyrmions move under
a drive and how they interact with a substrate potential or pinning.
In the absence 
of a substrate, the skyrmion moves at an angle with respect to the drive
which is known as the intrinsic skyrmion Hall 
angle $\theta_{sk}^{int}$ \cite{EverschorSitte18,Nagaosa13}.
The magnitude of this angle increases
as the ratio of the Magnus force to the
damping term increases.
In principle, skyrmion Hall angles 
of close to $\theta_{sk}=90^\circ$ are possible in certain systems.
Experimental measurements of $\theta_{sk}$ give
values ranging from just a few degrees up 
to $\theta_{sk}=55^\circ$ \cite{Jiang17,Litzius17,Woo18,Juge19,Zeissler19}.
These experiments are generally performed on
larger skyrmions that can be observed directly,
but much larger skyrmion Hall angles could be present
in systems with smaller skyrmions.
In race track devices, the skyrmion Hall angle can limit the distance the 
skyrmion can move,
so there is currently considerable interest in identifying ways to reduce or 
control the skyrmion Hall angle \cite{Fert17}.
The skyrmion
Hall angle is known to have a strong drive dependence
in the presence of pinning.
At low drives, the skyrmion Hall angle is small or zero, but $\theta_{sk}$ increases
with increasing drive before saturating at a value close to the pin-free limiting value
for high drives
\cite{Jiang17,Litzius17,Woo18,Juge19,Zeissler19,Reichhardt15,Legrand17,Reichhardt16,Kim17,Reichhardt19}. 
This drive dependence can arise due to a side jump or 
swirling motion of the skyrmions that occurs when they interact with
pinning sites, where faster moving skyrmions undergo a
smaller side jump \cite{Muller15,Reichhardt15a}.
Most work on skyrmions
and pinning has focused on systems with randomly placed pinning;
however, nanostructuring techniques provide
a wide variety of ways to create 
periodic substrates that
could act as attractive or repulsive scattering sites for skyrmions
\cite{Stosic17,Fernandes18}. 
Using a particle-based model for individual skyrmions in the presence of a 2D
pinning array,
Reichhardt {\it et al.}
demonstrated that the skyrmion Hall angle is initially small and increases with
increasing driving force, but that this increase takes the form of a series of
jumps, giving a quantization of the skyrmion Hall angle that is absent for
random pinning
\cite{Reichhardt15a}.
This is similar to the symmetry locking
found in superconducting vortex \cite{Reichhardt99,Reichhardt08a}
and colloidal systems \cite{Gopinathan04,Reichhardt04,MacDonald03} under a rotating
external drive; 
however, in the skyrmion system, the drive direction remains fixed and it is
the velocity dependence of the skyrmion Hall angle
which generates the change in the skyrmion flow direction. 
Individual skyrmions interacting with a periodic array of antidots 
have also been studied numerically
using micromagnetic and Theile equation approaches,
where it was shown
that the skyrmion motion can be carefully controlled
with a drive and
can lock to certain directions,
and that there are
localized states in which the skyrmion undergoes
circular motion in the
interstitial regions between antidots or moves around the edge of an antidot
\cite{Feilhauer19}.  

In this work we examine skyrmions interacting with a two dimensional periodic array of 
obstacles or antidots while being driven with a force that is applied in a
fixed direction.
We find directional symmetry locking of the
skyrmion motion similar to that reported in previous work \cite{Reichhardt15a};
however, here we also examine the effects of changing 
the obstacle size.  In some regimes
we find that the skyrmion motion locks parallel to the driving direction, and in
all cases the locking phases are associated with both a velocity locking of the
skyrmion and with a quantized skyrmion Hall angle. 
We also consider dc driving of
a mixture of skyrmion species with different Magnus forces through the
same periodic obstacle array.
In this case, the flow is more complex, the steps in the velocity-force
curves become less distinct, and we observe
directional locking effects in which one species
locks a symmetry direction while the other does not.   
This effect can be harnessed in order to achieve the
topological sorting of different skyrmion species
in analogy to the sorting of different species of particles or colloids moving
over periodic arrays in overdamped systems
\cite{Huang04,Ladavac04,Jonas08,Xiao10a,Tahir11,Risbud14}.
For increasing skyrmion density, the
sorting effect is slightly reduced but remains robust
up to very high skyrmion density, where a jammed phase appears in which
both species
move together either in the driving direction or
along a symmetry direction of the substrate array. 

\begin{figure}
  \begin{center}
    \includegraphics[width=0.8\columnwidth]{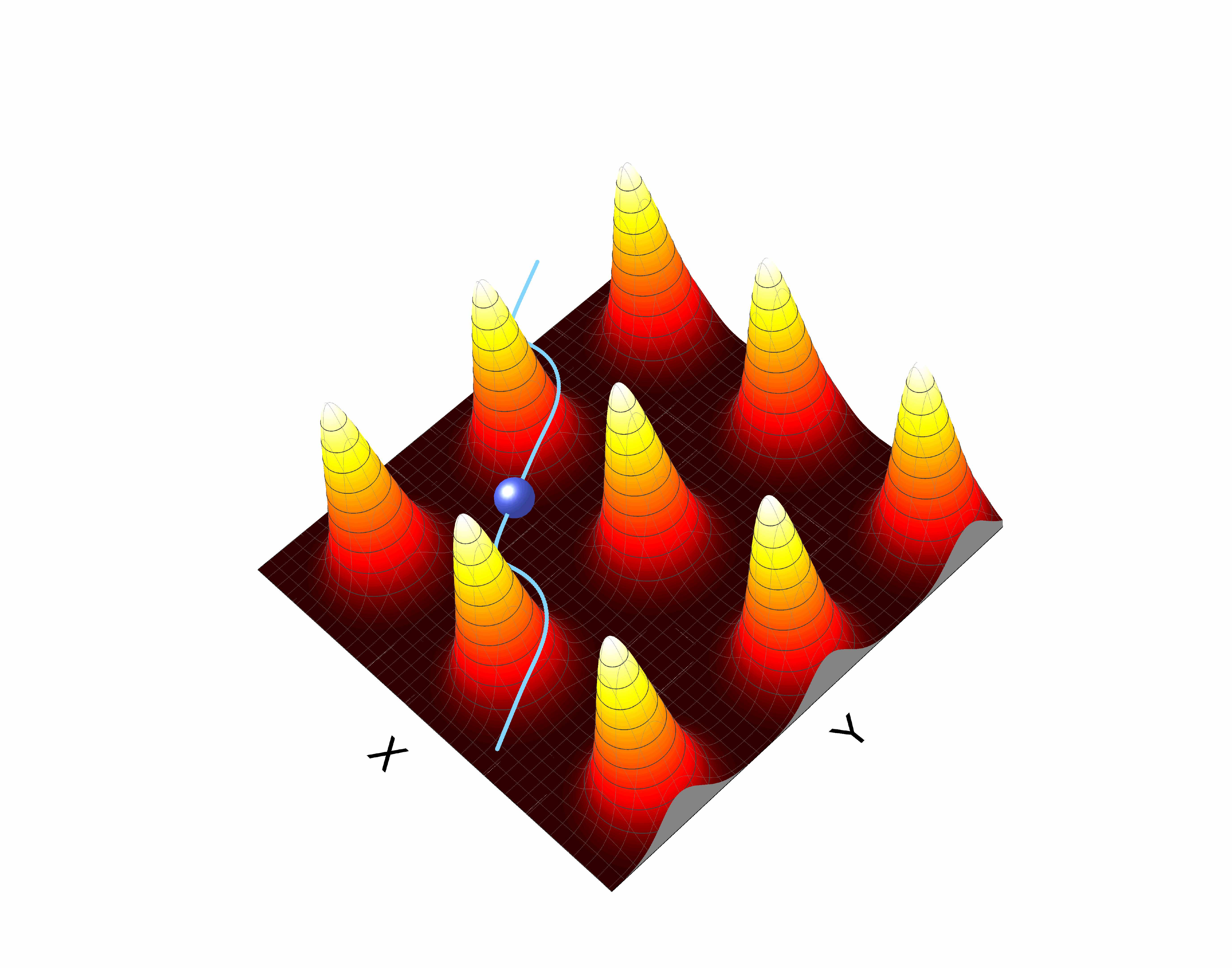}
    \end{center}
  \caption{Image of a portion of the sample showing a square array of obstacles
    (peaks) modeled as repulsive Gaussian scattering sites.  The blue bead represents
    a skyrmion, modeled as a point like particle with dynamics that contain both damping
    and a Magnus term.  The blue line indicates the path followed by the skyrmion under
    a dc drive applied along the $x$ direction.
}
\label{fig:1}
\end{figure}

\section{Simulation}

In this work we consider a two-dimensional skyrmion system of
size $L\times L$ with periodic boundary conditions in the $x$ and $y$ directions.
We initially focus on
a single skyrmion moving through a square obstacle array,
and later we introduce
multiple interacting
skyrmions of different types.
The total number of skyrmions is $N=N_a+N_b$, where $N_a$ ($N_b$)
is the number of skyrmions of species $a$ ($b$).
We set $N_a=N_b=N/2$ except when $N=1$.
The different skyrmion species
represent skyrmions with different sizes
that can coexist in a sample.
The total skyrmion density is $n_s=N/L^2$.

\begin{figure}
  \begin{center}
    \includegraphics[width=0.6\columnwidth]{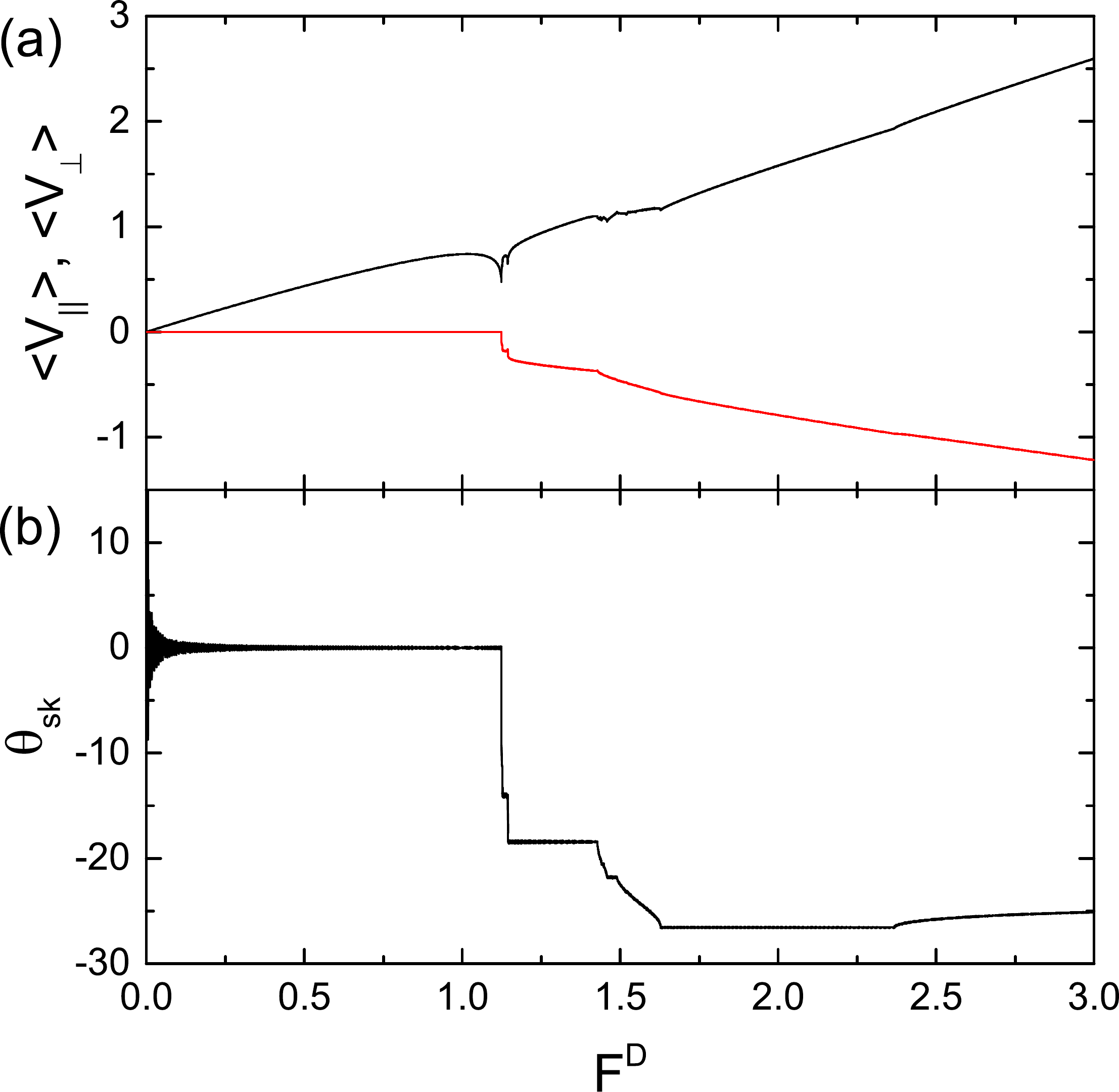}
    \end{center}
\caption{(a) $\langle V_{||}\rangle$ (black) and $\langle V_{\perp}\rangle$ (red)
  vs $F^{D}$ for a system containing a single skyrmion where
  $a_{0} = 0.65$ and $\alpha_{m}/\alpha_{d} = 0.45$.
  When $F^{D} < 1.0$, the motion is
  strictly along the $x$-direction, parallel to the drive.
  (b) The corresponding skyrmion Hall angle $\theta_{sk} = \arctan(R)$ vs $F^{D}$, 
  where $R = \langle V_{\perp}\rangle /\langle V_{||}\rangle.$
} 
\label{fig:2}
\end{figure}

\begin{figure}
  \begin{center}
    \includegraphics[width=0.6\columnwidth]{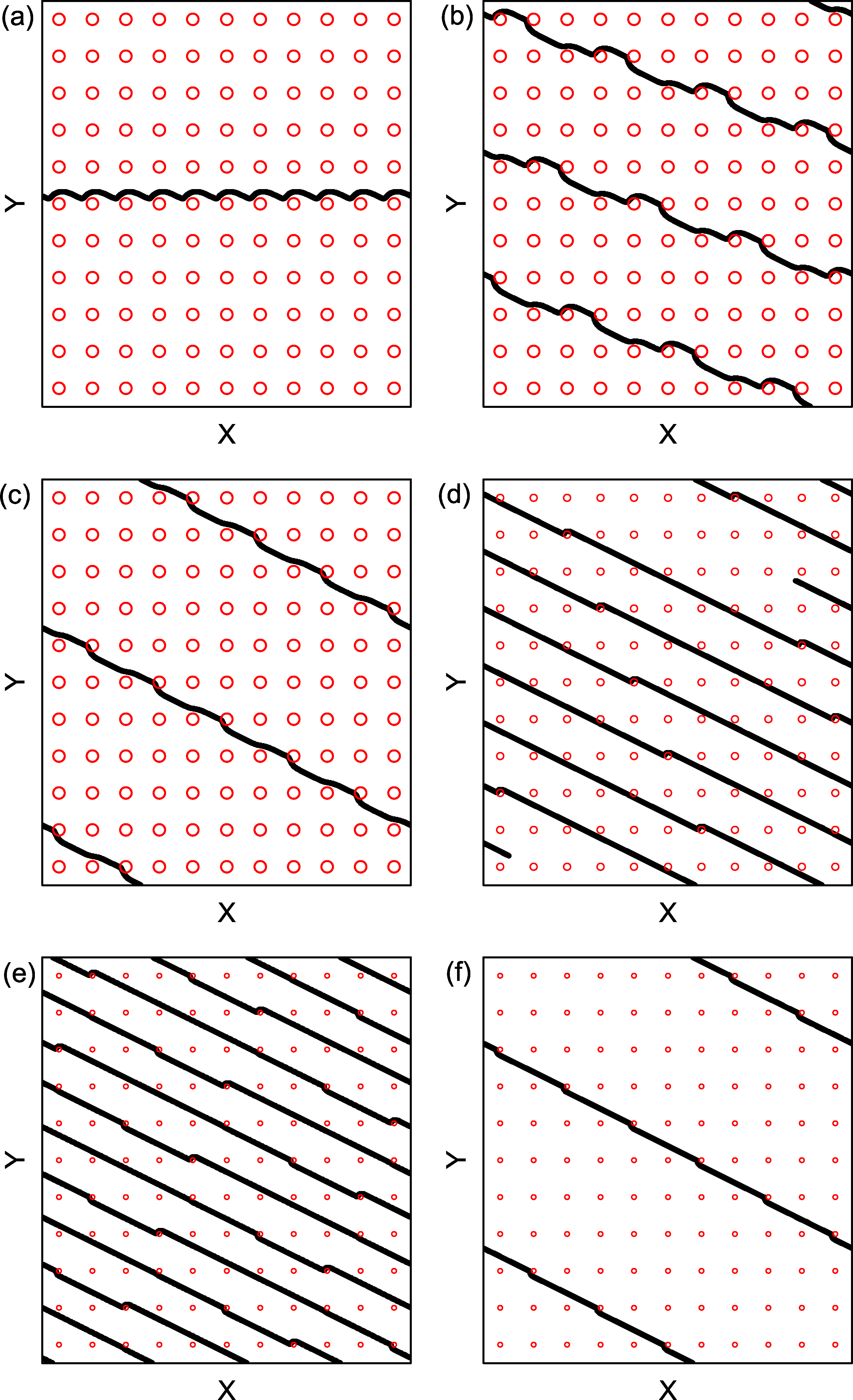}
    \end{center}
\caption{ (a-d) The obstacles (open circles) and the skyrmion trajectory (lines)
  for the system in Fig.~\ref{fig:2} with $a_0=0.65$
  and $\alpha_m/\alpha_d=0.45$. 
  (a) $F^{D} = 0.5$, where $R = 0.0$.
  (b) $F^{D} = 1.25$, where $R = -1/3$ and the system is 
  on a locking step.
  (c) $F^{D} = 2.0$, where $R = -0.5$.
  (d) Obstacles and skyrmion trajectories
  for the system in Fig.~\ref{fig:4} with $a_{0}=  0.15$ and
  $\alpha_m/\alpha_d=0.45$ at
  $F^{D} = 0.5$,
  where $R = -0.45$. 
  (e-f)  The same 
  for the system in Fig.~\ref{fig:4} with $a_{0} =  0.2$
  and $\alpha_m/\alpha_d=0.45$. 
  (e) $F^{D} = 0.5$, where $R = -0.5$. 
  (f) $F^{D} = 2.0$,
  where $R = -0.45$. }    
\label{fig:3}
\end{figure}

The dynamics of skyrmion $i$ is obtained using a
particle based skyrmion model \cite{Lin13},
with the following equation of motion:
\begin{equation}
 {\alpha }^{\gamma }_d{\rm {\bf{v}}}_i+{\alpha }^{\gamma }_m\hat{z}\times {\rm {\bf{v}}}_i={\rm {\bf{F}}}^{ss}_i+{\rm {\bf{F}}}^o_i+{\rm{\bf{F}}}^D .
\end{equation}
The first term on the left is the damping term for species $\gamma$,
where $\gamma=(a,b)$ and the damping constant is
${\alpha }^{\gamma }_d$. 
The damping originates from the spin precession and
dissipation of electrons localized in the skyrmion core. 
The second term on the left represents the Magnus force, 
where ${\alpha }^{\gamma }_d$ is the Magnus term for species $\gamma$. 
The Magnus force is oriented perpendicular to the skyrmion velocity. 
The repulsive skyrmion-skyrmion interaction force is given by
${\rm {\bf{F}}}^{ss}_i=\sum^N_i{K_1\left(r_{ij}/\xi \right){\widehat{\rm {\bf{r}}}}_{ij}}$.
Here, $\xi$ is the screening length which we take to be $1.0$ in dimensionless units,
$r_{ij}=\left|{\rm {\bf{r}}}_i-{\rm {\bf{r}}}_j\right|$ is the
distance between skyrmions $i$ and $j$, and
${\widehat{\rm {\bf{r}}}}_{ij}=\left({\rm {\bf{r}}}_i-{\rm {\bf{r}}}_j\right)/r_{ij}$.
In order to enhance computational efficiency,
we cut off the skyrmion-skyrmion interaction beyond $r_{ij}=6.0$ where its magnitude
becomes negligible.
We set the sample size to $L=36\xi$.
We model the potential energy of the skyrmion-obstacle interaction using
the Gaussian form $U_o=C_oe^{-{\left({r_{io}}/{a_o}\right)}^2}$, where $C_o$ 
is the strength of the obstacle potential. 
The skyrmion-obstacle interaction is thus given by
${\rm {\bf{F}}}^o_i=-\mathrm{\nabla }U_o=-F_or_{io}e^{-{\left({r_{io}}/{a_o}\right)}^2}{\widehat{\rm {\bf {r}}}}_{io}\ $,
where $F_o=2U_o/a^2_o$, 
$r_{io}$ is the distance between skyrmion $i$ and obstacle $o$,
and $a_o$ is the obstacle radius. 
We cut off this interaction beyond
$r_{io}=2.0$ where the force becomes negligible.
We consider an obstacle density of
$\rho_o=0.093/{\xi }^2$.
The skyrmion-external current interaction is given by
${\rm {\bf {F}}}^D =F^D\widehat{\rm {\bf{d}}}$,
where $\widehat{\rm {\bf{d}}}$ is the direction of the driving force,
which is fixed in the $x$ direction,
$\widehat{\rm {\bf{d}}}=\ \widehat{\rm {\bf{x}}}$.

In Fig.~\ref{fig:1} we show a schematic of our system highlighting the obstacle array 
and a skyrmion moving through the array.  
We measure the
skyrmion velocity parallel, $\left\langle V_{\parallel }\right\rangle $, and perpendicular, $\left\langle V_{\bot }\right\rangle $, to the drive.
When the skyrmion is flowing in the absence of obstacles,
in the overdamped limit of ${\alpha }_m/{\alpha }_d=0$
it moves only in the
direction of the drive.
If ${\alpha }_m/{\alpha }_d$ is finite, the skyrmion
instead moves with a Hall angle of
$\theta_{sk}={\mathrm{arctan} \left(\left\langle V_{\bot }\right\rangle /\left\langle V_{\parallel }\right\rangle \right)\ }={\mathrm{arctan} \left({\alpha }_m/{\alpha }_d\right)\ }$.
In order to quantify the direction of the skyrmion motion we measure
$R=\ \left\langle V_{\bot }\right\rangle /\left\langle V_{\parallel }\right\rangle$, where the
skyrmion Hall angle is given by $\theta_{sk} = \arctan(R)$.
We increase the drive in small steps of $\delta F=0.001$ and
we wait ${10}^5$ simulation time steps between increments
to ensure that the system has reached a steady state.
Unless otherwise noted, we normalize the
damping and Magnus coefficients such that $({\alpha^\gamma })^2_d+({\alpha^\gamma })^2_m=1$.

\begin{figure}
  \begin{center}
    \includegraphics[width=0.6\columnwidth]{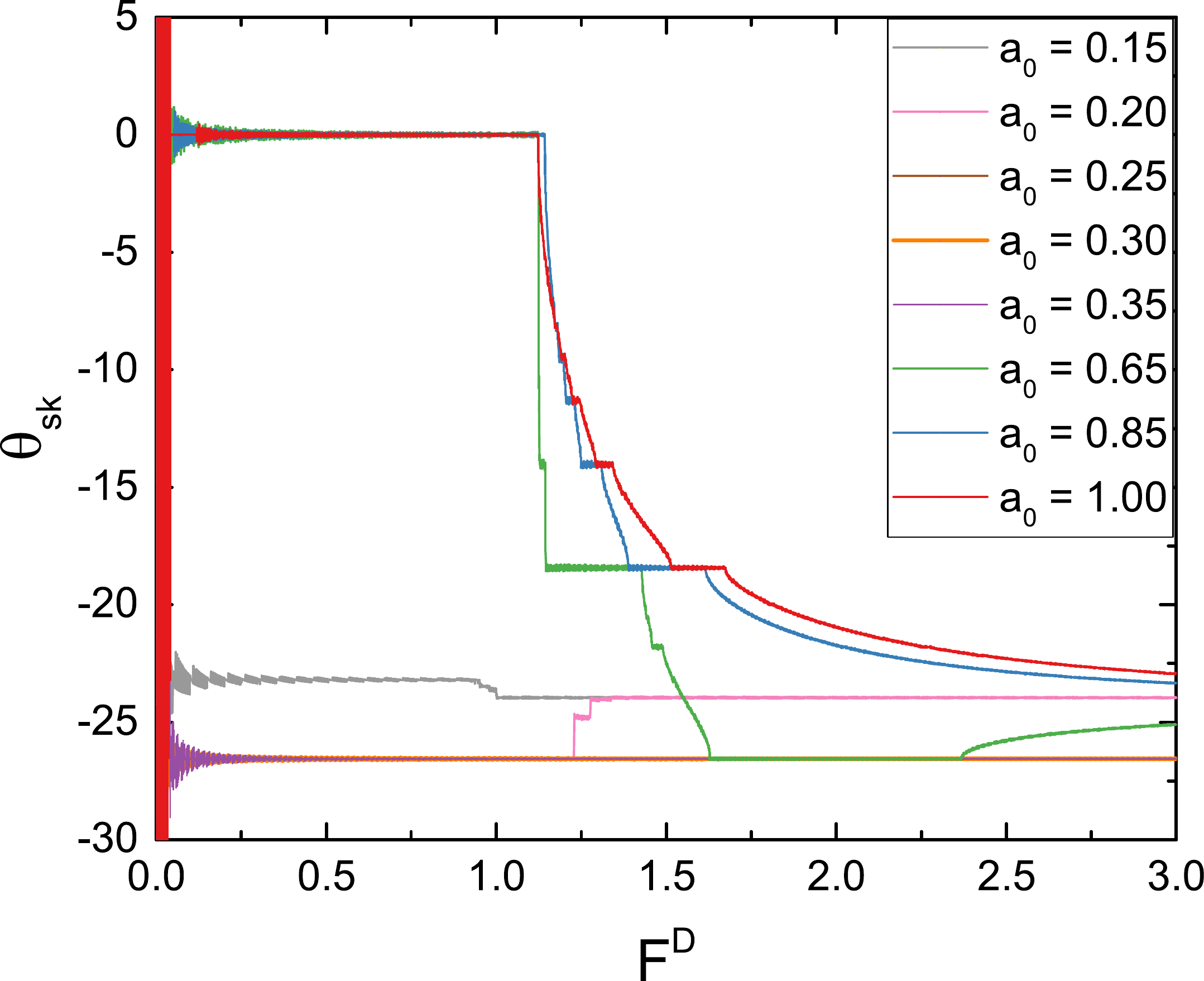}
    \end{center}
\caption{The skyrmion Hall angle $\theta_{sk} = \arctan(R)$ vs $F^{D}$,
  where $R = \langle V_{\perp}\rangle/\langle V_{||}\rangle$,
  in samples with $\alpha_{m}/\alpha_{d} = 0.45$
  at varied $a_{0} = 0.15$, 0.2, 0.25, 0.3, 0.35, 0.65, 0.85, and $1.0$.}
\label{fig:4}
\end{figure}

\section{Symmetry Locking and Quantized Hall Angle} 

\subsection{Changing obstacle size }

We first consider the case of a single skyrmion moving through 
an obstacle array for varied obstacle size and driving force.  
In Fig.~\ref{fig:2}(a) we plot $\langle V_{||}\rangle$ and
$\langle V_{\perp}\rangle$ versus $F^{D}$ for a system with fixed 
$a_{0} = 0.65$ and $\alpha_{m}/\alpha_{d} = 0.45$, while
Figure~\ref{fig:2}(b) shows the
corresponding skyrmion Hall angle 
$\theta_{sk} = \arctan(R)$.
If there are no obstacles,
the skyrmions move at an intrinsic angle of $\theta_{sk}^{\rm int} = 24.23^\circ$.  
We find that when $F^{D} < 1.0$,
$\langle V_{\perp}\rangle= 0$ and $\langle V_{||}\rangle$ increases with
increasing $F^D$, so that the skyrmion Hall angle is
$\theta_{sk} = 0$. 
As shown in Fig.~\ref{fig:3}(a), at $F^{D} = 0.5$ 
the skyrmion exhibits an oscillatory motion but translates only in
the $x$-direction. 
As $F^{D}$ increases,
the skyrmion starts to move in both the $+x$ and $-y$ directions.
At the transition
to finite $\langle V_{\perp}\rangle$,
there is a drop in $\langle V_{||}\rangle$, indicating that the particle
is actually slowing down in the $x$-direction
as a function of
increasing $F^{D}$, indicating the appearance of
negative differential conductivity
with $d\langle V_{||}\rangle/dF^D < 0.0$.
For higher $F^D$,
several
additional dips or cusps appear in both $\langle V_{||}\rangle$
and $\langle V_{\perp}\rangle$.
As shown in Fig.~\ref{fig:2}(b),
the skyrmion Hall angle is quantized,
with steps at
$R=-0.33$ ($\theta_{sk} = -18.43^\circ$) and
$R = -0.25$ ($\theta_{sk} = -14.03^\circ)$,
corresponding to orbits in which the skyrmion moves one lattice
constant in the $-y$-direction
and three or four lattice constants in the $x$-direction, respectively. 

In Fig.~\ref{fig:3}(b), we illustrate the skyrmion orbit for
the $R = -0.33$ step at $F^{D} = 1.25$. 
The orbit winds periodically around the system and the skyrmion undergoes
multiple collisions with the
obstacles during each complete orbit.
As $F^D$ increases,
the
magnitude of both $\langle V_{||}\rangle$ and $\langle V_{\perp}\rangle$
jumps up and the system reaches the
$R = -0.5$ step.
The corresponding orbit appears in Fig.~\ref{fig:3}(c) for $F^{D} = 2.0$,
where the skyrmion moves 2 lattice constants in
the $+x$ direction for every one lattice constant in the 
$-y$ direction. 
Interestingly, this means
that the skyrmion is moving at an angle of $\theta_{sk} = -26.565^\circ$, which 
is {\it larger} than the intrinsic skyrmion Hall angle
of $\theta_{sk}^{int}=-24.23^\circ$.
The strong locking on the $R=-0.5$ step
pulls the skyrmion to this higher value of $\theta_{sk}$ over
an interval of $F^D$ due to the symmetry of the square obstacle lattice.
The skyrmion remains
locked to the $R = -0.5$ state until $F^{D} > 2.3$,
after which the skyrmion motion drops to
a lower value of $\theta_{sk}$, which
rapidly approaches the intrinsic skyrmion Hall angle value.

\begin{figure}
  \begin{center}
    \includegraphics[width=0.6\columnwidth]{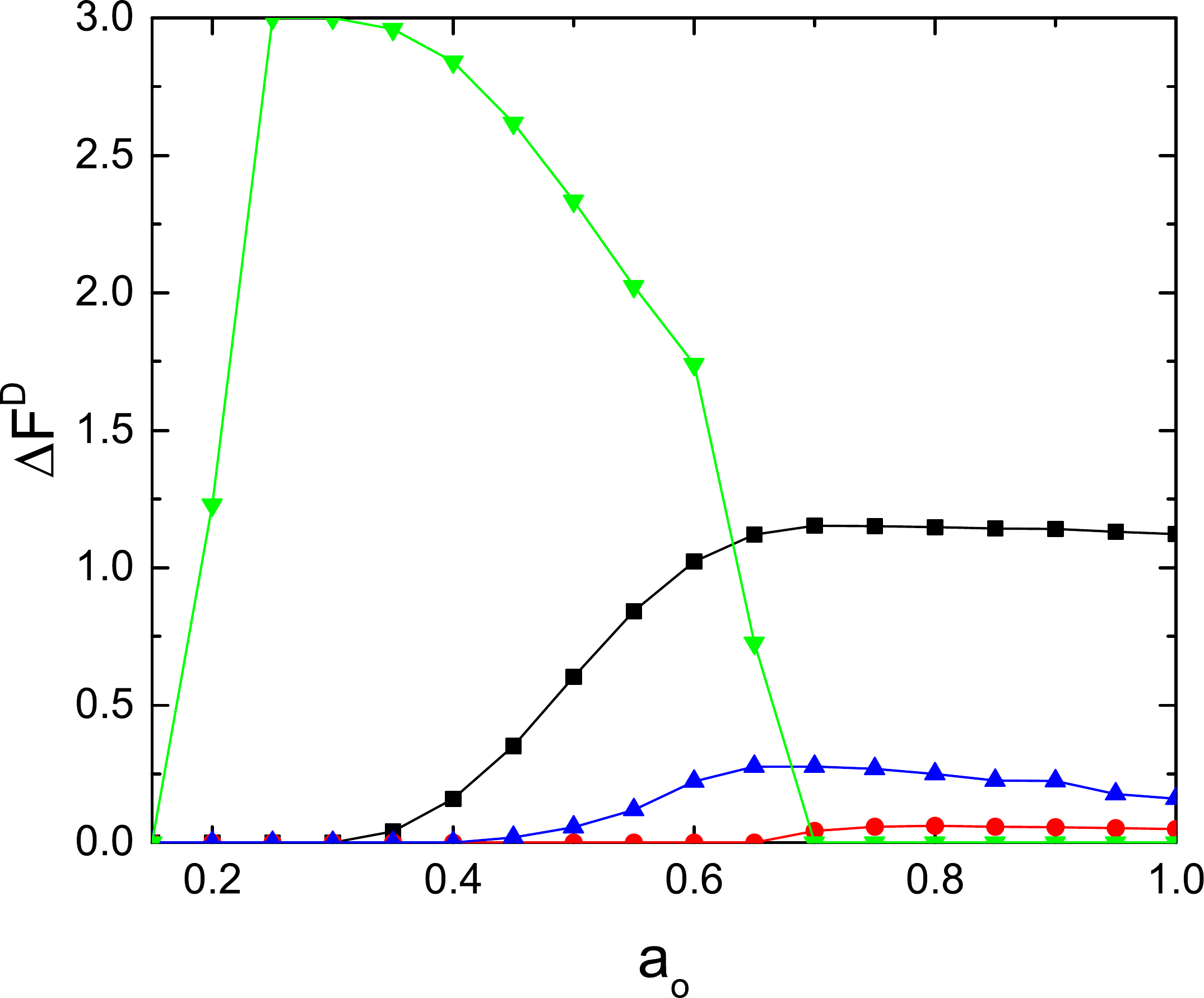}
    \end{center}
\caption{ The locking step force interval $\Delta F^{D}$
  vs $a_{0}$ at $\alpha_{m}/\alpha_{d} = 0.45$
  for the $R = 0.0$ (black squares), -0.25 (red circles), -0.33 (blue up triangles),
  and $-0.5$ (green down triangles) locking steps.   
 }
\label{fig:5}
\end{figure}

In Fig.~\ref{fig:4} we plot $\theta_{sk}$ versus $F^{D}$
for a system with $\alpha_{m}/\alpha_{d} = 0.45$ at varied
$a_{0} = 0.15$ to $1.0$.
For smaller $a_{0}$, the skyrmion initially moves at
an angle close to the
intrinsic skyrmion Hall angle of $\theta_{sk}^{int}=-24.23^\circ$ since
it experiences very few collisions with the obstacles.
When $a_0 = 0.2$, the
system starts off in the $R = -0.5$ ($\theta_{sk}=-26.565^\circ$)  state and jumps to  
the $R = -0.45$ ($\theta_{sk}=24.23^\circ$) state at higher drives.
For $a_{0} \geq 0.65$, the system
initially locks to the $R= 0.0$ ($\theta_{sk}=0$) state.
At the largest value of $a_0 = 1.0$, we find a series
of smaller steps at rational fractional ratios of
$R = -n/m$, and at large drives $R$ gradually approaches the
intrinsic value.
In Fig.~\ref{fig:3}(d) we plot the skyrmion trajectory
for the system in Fig.~\ref{fig:4} with $a_0=0.15$ at
$F^{D} = 0.5$,
where the motion is locked to the $R  = -0.45$ state,
while in Fig.~\ref{fig:3}(e)
at $F^D=0.5$ and $a_{0} = 0.20$,
the motion is now locked to the $R = -0.5$ state, which has a distinct trajectory.
Figure~\ref{fig:3}(f) shows that at
$F^{D} = 2.0$ and $a_{0} = 0.2$, the higher drive
$R = -0.45$ state has
a trajectory that is closer to linear.

We can also characterize the different locking phases by examining the
width of the force interval 
$\Delta F^D$ over which the system remains
locked to a given step for varied $a_{0}$ in samples
with $\alpha_{m}/\alpha_{d} = 0.45$.
In Fig.~\ref{fig:5}  we plot
the locking intervals $F^D$ for 
the $R = 0.0,$ -0.25, -0.33, and $-0.5$ steps.
The $R=0.0$ locking step 
appears when $a_{0} > 0.3$
and saturates to a maximum width near $a_{0} = 0.7$.
The width of the $R = -0.5$ step is large when $a_0=0.25$, decreases 
for increasing $a_{0}$, and drops to zero for $a_{0} > 0.7$.
The widths of the $R = -0.25$ and $R=-0.5$ steps
are smaller but generally increase with increasing $a_{0}$.
We note that behavior 
similar to that shown for the $R = -0.25$ locking step
appears at the higher order values of $R$ such as $1/5$ and $1/6$.

\begin{figure}
  \begin{center}
    \includegraphics[width=0.9\columnwidth]{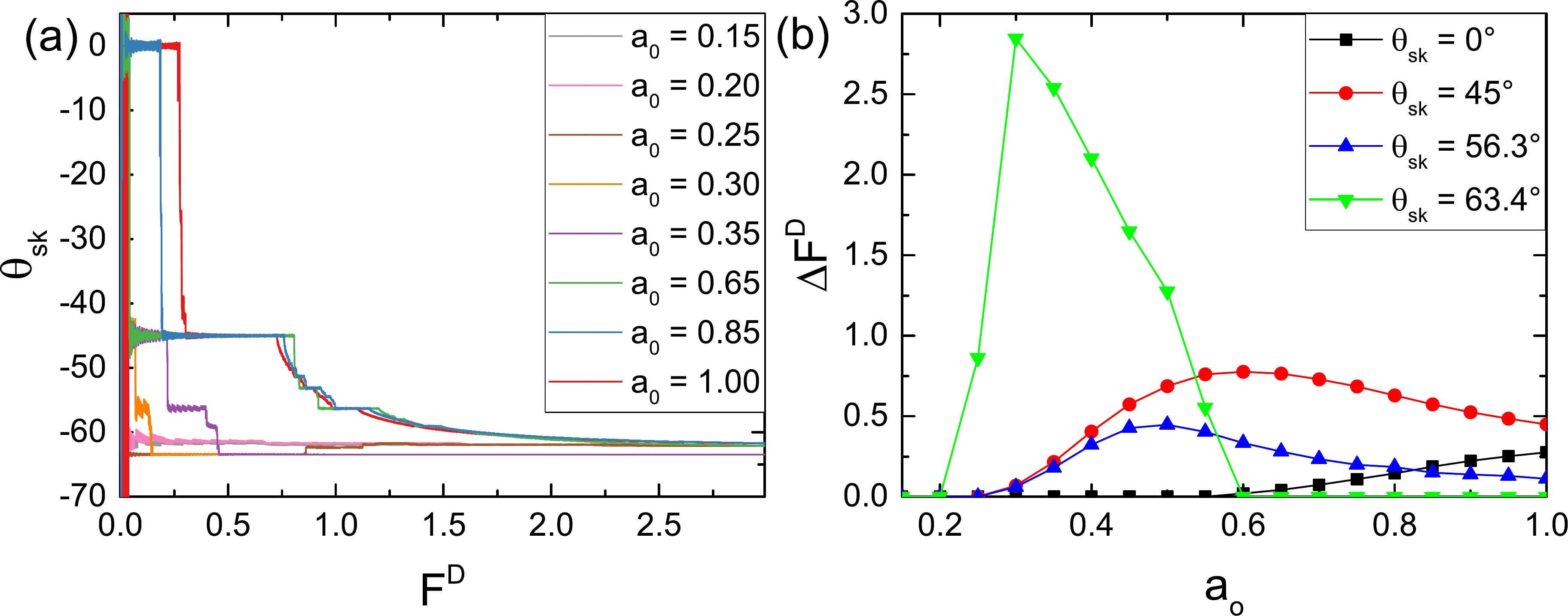}
    \end{center}
  \caption{(a) $\theta_{sk} = \arctan(R)$  vs $F^{D}$ for varied $a_{0}$
    of 0.15, 0.20, 0.25, 0.30, 0.35, 0.65, 0.85, and 1.0
    in samples with $\alpha_{m}/\alpha_{d} = 1.91$.
    Locking steps appear at
    $R = 0$, -1.0, -1.5, -1.91, and $-2.0$.  
    (b) $\Delta F^{D}$ vs $a_{0}$ for the system in (a)
    for the $R = 0$ (black squares), -1.0 (red circles), -1.5, (blue up triangles),
    and $2.0$ (green down triangles) steps.
}
\label{fig:6}
\end{figure}

\begin{figure}
  \begin{center}
    \includegraphics[width=0.6\columnwidth]{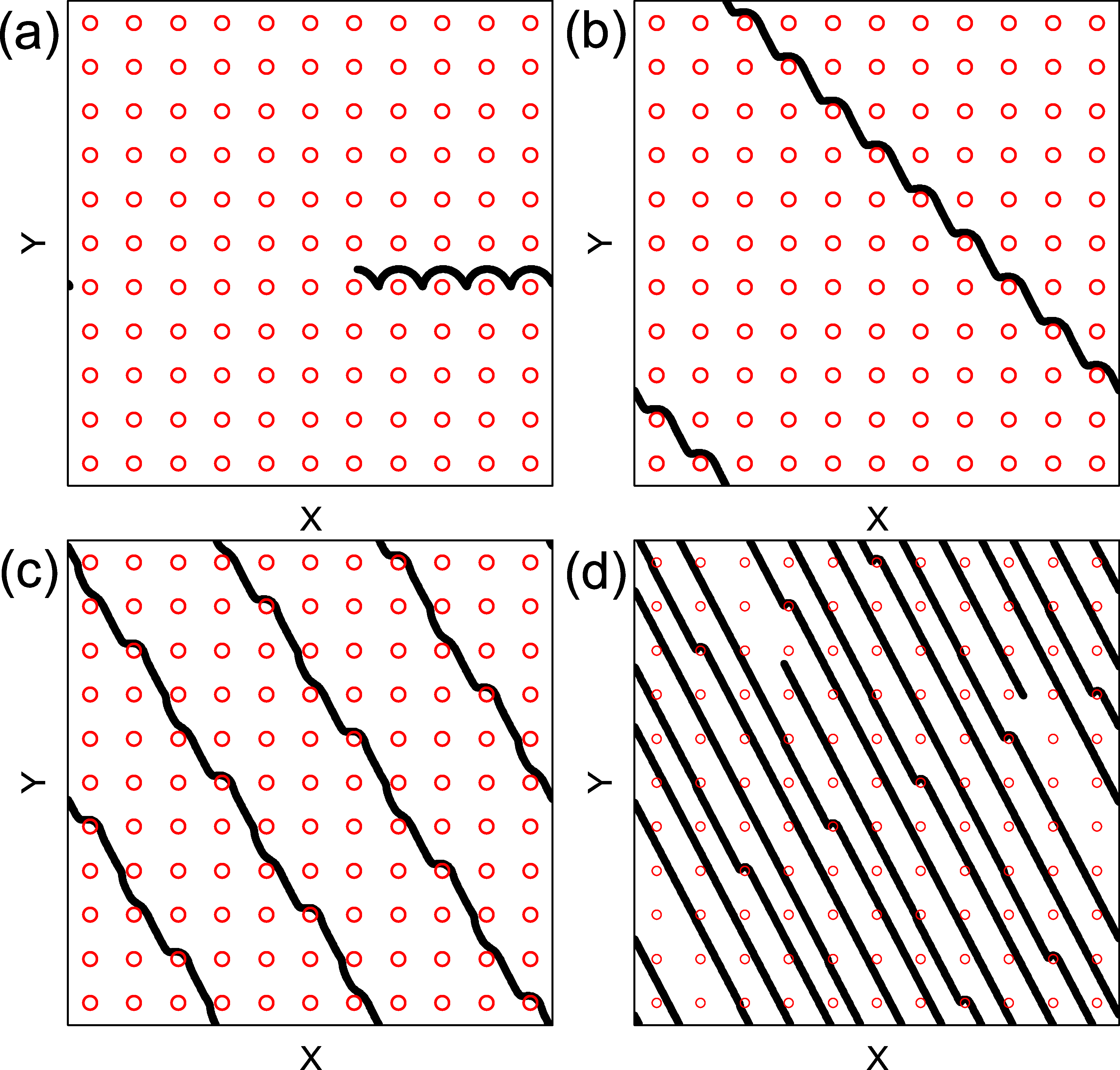}
    \end{center}
\caption{
  The obstacles (open circles) and the skyrmion trajectory (lines)
  for the system in Fig.~\ref{fig:6} with
  $\alpha_{m}/\alpha_{d} = 1.91$.
  (a)  The $R = 0.0$ state for $a_0=0.65$ at $F^{D} = 0.04$.
  (b) The $R = -1.0$ state for $a_0=0.65$ at $F^{D} = -1.0$.
  (c) The $R = -1.5$ state for $a_0=0.65$ at $F^{D} = 1.0$.
  (d) The $R=-1.85$ state for $a_0=0.2$ at $F^D=1.0$.
}
\label{fig:7}
\end{figure}

For increased $\alpha_{m}/\alpha_{d}$, the skyrmion can move at larger Hall angles,
making it possible to observe a larger number of
locking phases.
In Fig.~\ref{fig:6}(a) 
we plot $\theta_{sk}$ versus $F^{D}$ 
for samples with $\alpha_{m}/\alpha_{d} = 1.91$ at $a_0=0.15$ to 1.0.
We find steps at  
$R = -1.0$, -1.5, -2.0, and $-1.91$,
corresponding to $\theta_{sk}=-45^\circ$, $-56.3^\circ$, $-63.43^\circ$,
and $-62.37^\circ$, respectively,
along with numerous higher order steps.
When $a_{0} = 0.15$,
the skyrmion remains locked
at $ R = -1.91$, while when $a_{0} = 0.25$,
the system is initially locked to $R = -2.0$, corresponding to a Hall angle that
is higher than the intrinsic $\theta_{sk}^{int}$, and
for higher drives it undergoes
two transitions before reaching $R = -1.91$.
When $a_{0} > 0.64$, the  $R = 0$ state appears at low drives,
and for $a_{0} = 0.85$,
there are a number of additional steps
corresponding to
$R = -1.25$, -1.33, and  $-1.67$, which are equivalent to
$\theta_{sk}=-51.34^\circ$, $53.06^\circ$, and $-59.09^\circ$, respectively.
In Fig.~\ref{fig:7}(a) we plot the skyrmion trajectories for the system in 
Fig.~\ref{fig:6} with $a_{0} = 0.65$ at $F^{D} = 0.04$ 
in the $R = 0.0$ state, while
in Fig.~\ref{fig:7}(b) we show the trajectories
in the same system for the $R = -1.0$ state at $F^{D} = 0.5$, where the skyrmion
moves at an angle of  $\theta_{sk}=-45^\circ$. 
At $F^D=1.0$ in the $R=-1.5$ state, illustrated
in Fig.~\ref{fig:7}(c),
the skyrmion Hall angle is
$\theta_{sk} = 56.31^\circ$ and
the skyrmion moves $3a$ in the $y$-direction
while translating a distance $2a$ in the $x$-direction.
If the pinning site radius is reduced to $a_0=0.2$,
then for $F^D=1.0$ the system is in the $R=-1.85$ state, shown
in Fig.~\ref{fig:7}(d).
In Fig.~\ref{fig:6}(b), we plot $\Delta F^{D}$ vs $a_{0}$ for
the same system highlighting the regions in which the
$R = 0$, -1, -1.5, and $-2.0$ states appear.
We find that the widths of the $R = -1.0$ and $R=-1.5$ states
are maximized at a particular value of $a_{0}$.

\begin{figure}
  \begin{center}
    \includegraphics[width=0.9\columnwidth]{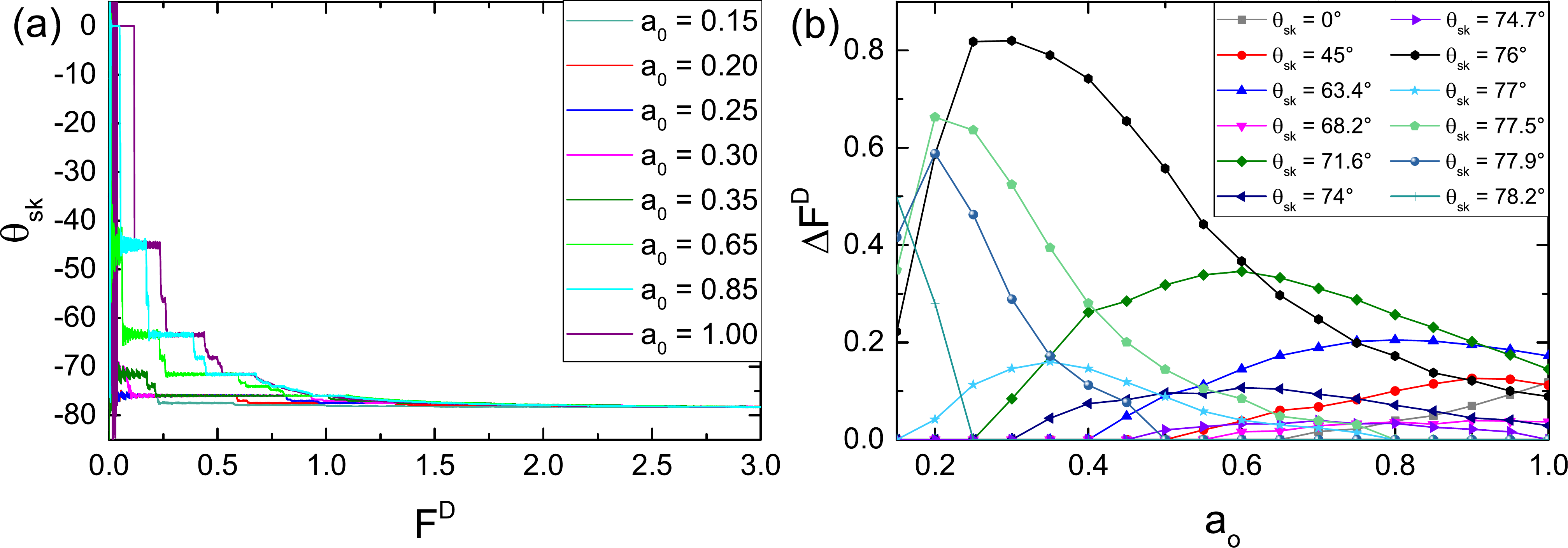}
    \end{center}
\caption{ 
  (a) $\theta_{sk} = \arctan(R)$  vs $F^{D}$ for varied $a_{0}$ of
  0.15, 0.20, 0.25, 0.30, 0.35, 0.65, 0.85, and 1.0
  in samples with $\alpha_{m}/\alpha_{d} = 4.925$.
  (b) $\Delta F^{D}$ vs $a_{0}$ for the system in (a) for
  $R = 0.0$ to $R= -4.8$.  As $a_0$ increases, a larger
  number of steps appear.
}
\label{fig:8}
\end{figure}

\begin{figure}
  \begin{center}
    \includegraphics[width=0.6\columnwidth]{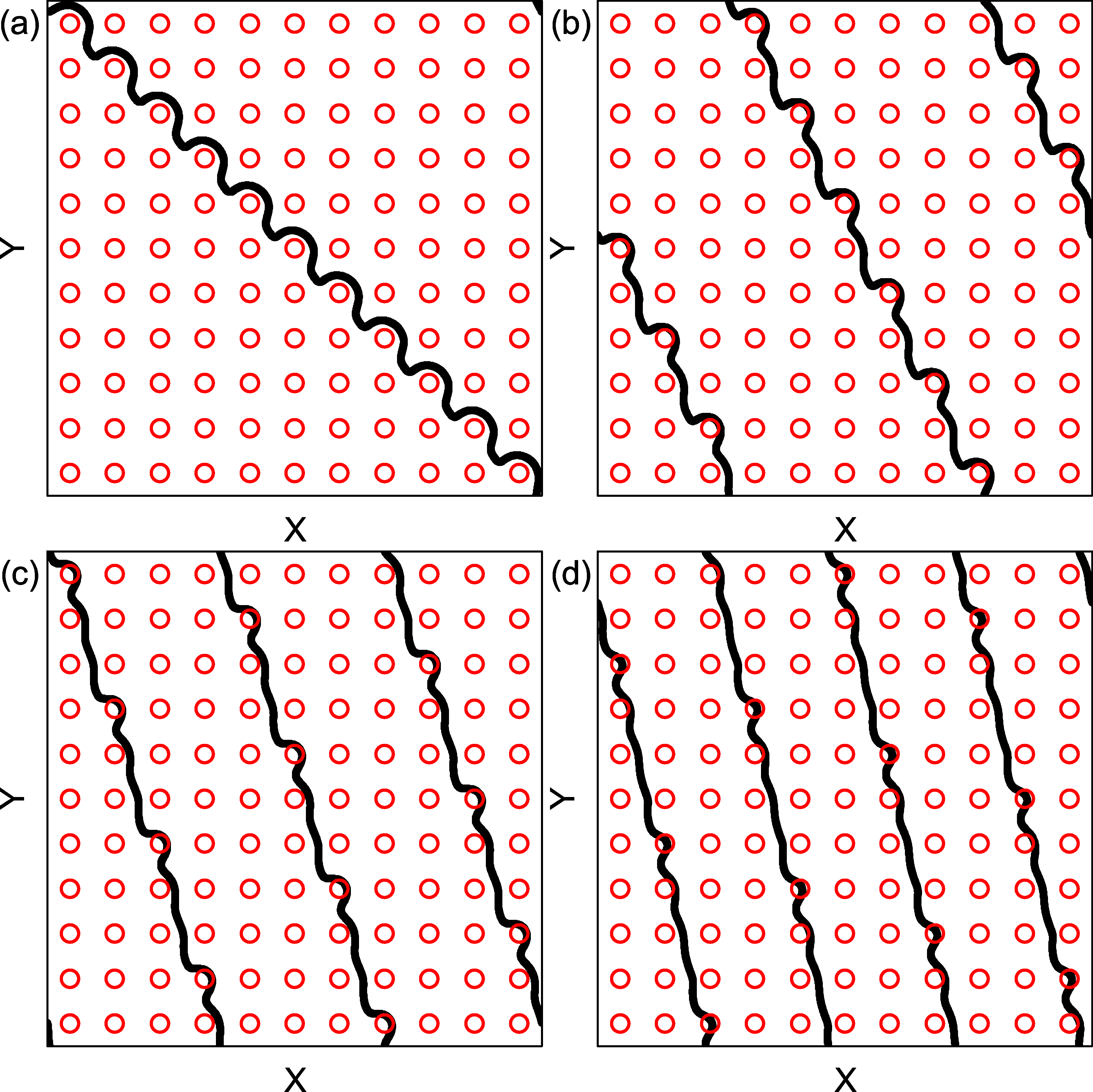}
    \end{center}
\caption{ The obstacles (open circles) and the skyrmion trajectory (lines)
  for the system in Fig.~\ref{fig:8} with
  $\alpha_m/\alpha_d=4.925$ at $a_{0} = 1.0$.
  (a) $R = -1.0$. (b) $R = -2.0$.
(c) $R = -3.0$. (d) $R = -4.0$. 
}
\label{fig:9}
\end{figure}

In Fig.~\ref{fig:8}(a) we plot $R$ versus $F^{D}$ for
samples with $\alpha_{m}/\alpha_{d} = 4.925$ where
$a_{0}$ ranges from $a_0=0.15$ to $a_0=1.0$.
Here we find
a larger number of possible locking steps
ranging from $R = 0.0$ to $R = -4.8$,
with prominent steps appearing
at $R = 0.0$, -1, -2, -3, and  $-4$
as well as numerous additional smaller steps such as those at
$R=-2.5$ or $R=-3.5$.
In Fig.~\ref{fig:8}(b) we show
$\Delta F^D$ versus $a_{0}$ for the $R=0$ to $R=-4.8$ steps
for the system in Fig.~\ref{fig:8}, where
the number of steps present increases as $a_0$ increases.
In Fig.~\ref{fig:9} we illustrate some of
the skyrmion trajectories that appear at
$a_{0} = 1.0$.
Figure~\ref{fig:9}(a) shows the $R = -1.0$ state
where the skyrmion follows a sinusoidal path
between the obstacles.
In Fig.~\ref{fig:9}(b) we plot the trajectories for the $R = -2.0$ state,
while in Fig.~\ref{fig:9}(c) we show the $R = -3.0$ state.
At $R=-4.0$ in Fig.~\ref{fig:9}(d),
the skyrmion trajectory becomes increasingly
tilted.  

\subsection{Changing Damping Effect}

\begin{figure}
  \begin{center}
    \includegraphics[width=0.6\columnwidth]{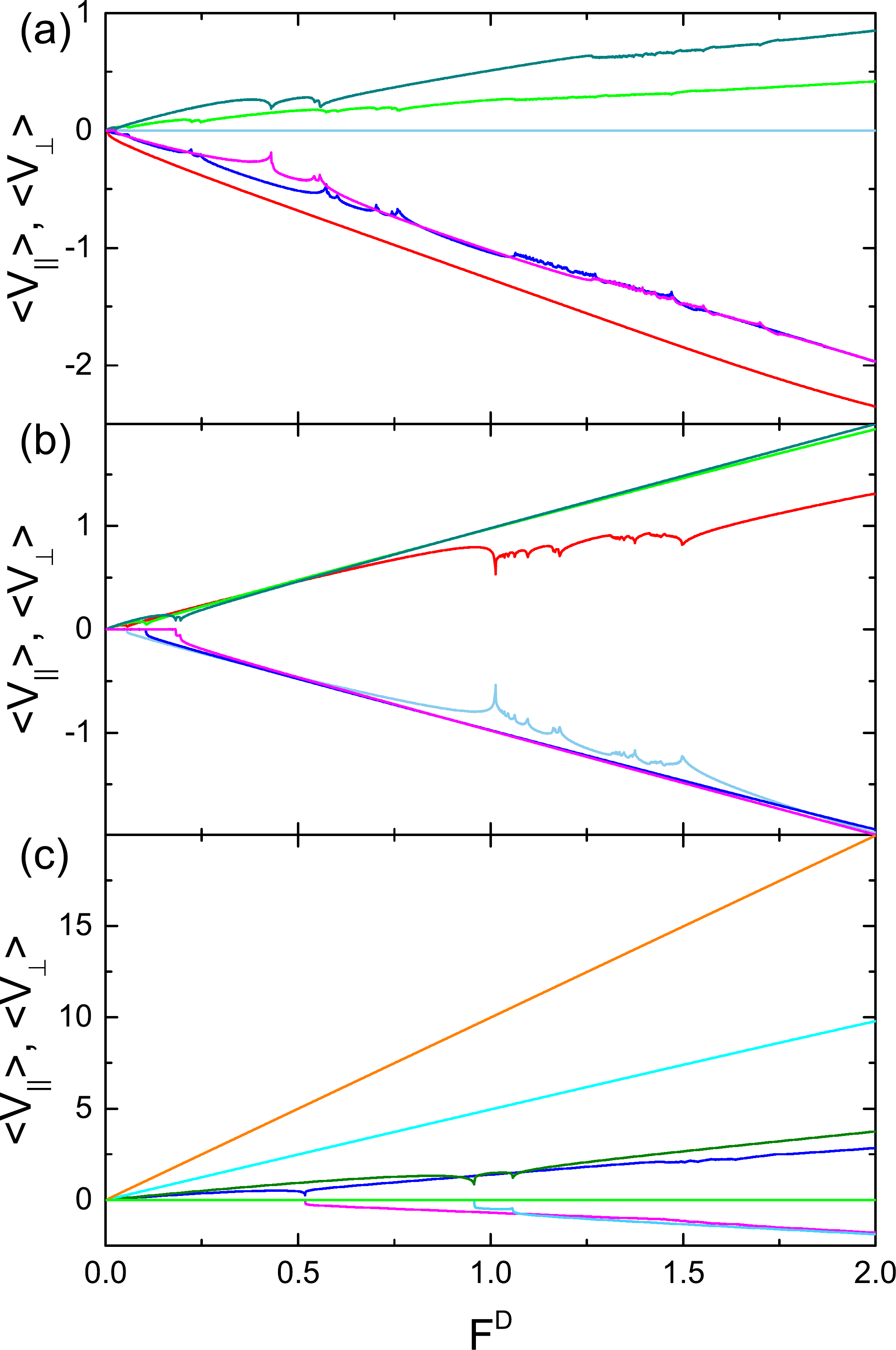}
  \end{center}
\caption{ (a) $\langle V_{||}\rangle$ (curves above zero) and
  $\langle V_{\perp}\rangle$ (curves below zero)
  vs $F^{D}$ for a system with 
  $\alpha_{m} = 1.0$, $a_{0} = 0.65$, and varied $\alpha_{d}$.
  (a) $\alpha_d= 0.0$
  (light blue, $\langle V_{||}\rangle$;
  red, $\langle V_{\perp}\rangle$),
  $\alpha_d=0.2$
  (light green, $\langle V_{||}\rangle$;
  dark blue, $\langle V_{\perp}\rangle$),
  and $\alpha_d=0.4$
  (dark green, $\langle V_{||}\rangle$;
  magenta, $\langle V_{\perp}\rangle$).
  (b) $\alpha_{d} = 0.6$
  (red, $\langle V_{||}\rangle$;
  light blue, $\langle V_{\perp}\rangle$);
  $\alpha_d=0.8$
  (light green, $\langle V_{||}\rangle$;
  dark blue, $\langle V_{\perp}\rangle$);
  and $\alpha_d=1.0$
  (dark green, $\langle V_{||}\rangle$;
  magenta, $\langle V_{\perp}\rangle$).
  (c)  $\alpha_{d} = 1.5$
  (dark blue, $\langle V_{||}\rangle$;
  magenta, $\langle V_{\perp}\rangle$);
  $\alpha_d=2.0$
  (dark green, $\langle V_{||}\rangle$;
  light blue, $\langle V_{\perp}\rangle$);
  $\alpha_d=5.0$
  (cyan, $\langle V_{||}\rangle$;
  red, $\langle V_{\perp}\rangle$);
  and $\alpha_d=10.0$
  (orange, $\langle V_{||}\rangle$;
  light green, $\langle V_{\perp}\rangle$).
}
\label{fig:10}
\end{figure}

\begin{figure}
  \begin{center}
    \includegraphics[width=0.6\columnwidth]{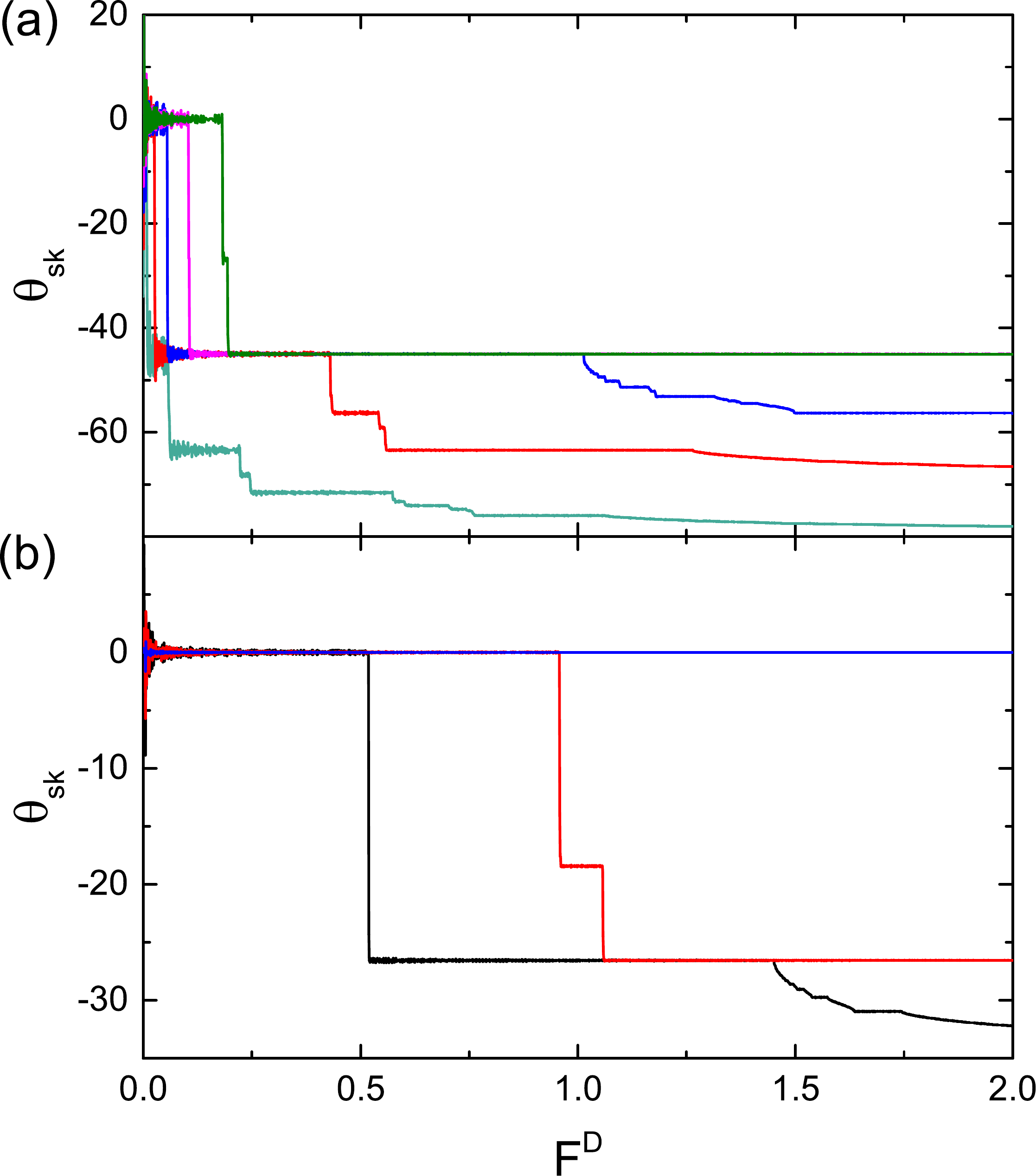}
    \end{center}
\caption{ $\theta_{sk}$ vs $F^{D}$ for the system in Fig.~\ref{fig:10}(a,b)
  with $\alpha_m=1.0$ and $a_0=0.65$ at different values of $\alpha_d$.
  (a) $\alpha_{d} = 0.2$ (teal),
  0.4 (red),
  0.6 (blue),
  0.8 (magenta),
  and $1.0$ (green). 
  (b)
  $\alpha_d=1.5$ (black), 2.0 (red), and 5.0 (blue).
  For $\alpha_{d} = 5.0$ and higher the 
  system remains locked to $\theta_{sk} = 0.0^\circ$.
}
\label{fig:11}
\end{figure}

We next consider the effect of changing the damping 
by fixing $\alpha_{m} = 1.0$ and varying $\alpha_{d}$.
In Fig.~\ref{fig:10}(a) we plot the perpendicular
$\langle V_{\perp}\rangle$ and parallel
$\langle V_{||}\rangle$ velocity components versus $F^D$ for a 
system with $a_{0} = 0.65$ at
$\alpha_d = 0.0$, 0.2, and $0.4$.
We observe
a series of dips in both
velocity components that are associated with the directional locking effects.
In Fig.~\ref{fig:10}(b),
where $\alpha_{d} = 0.6$, 0.8, and $1.0$, the 
$\alpha_{d} = 0.6$ curves exhibit the most prominent steps. 
The velocity force curves for $\alpha_{d} = 1.5$, 2.0, 5.0, and $10.0$
in Fig.~\ref{fig:11}(b) indicate that
$\langle V_{||}\rangle$ becomes more dominant as $\alpha_d$ increases,
and we find that
at $\alpha_{d} = 5$ and above, the system 
remains locked in the $R = 0.0$ state over the entire range of $F^D$.

In Fig.~\ref{fig:11}(a) we plot $\theta_{sk}$ versus $F^{D}$ for the system in
Fig.~\ref{fig:10}(a,b) with $\alpha_{d} = 0.2$, 0.4, 0.6, 0.7, and $1.0$.
As $\alpha_d$ increases,
fewer steps appear in $\theta_{sk}$.
The largest number of transitions occur
for the lowest damping of $\alpha_{d} = 0.2$,
and in all cases, we find a locking step corresponding to
$R = -1.0$.
Figure~\ref{fig:11}(b) shows the $\theta_{sk}$ versus $F^D$ curves for
$\alpha_{d} = 1.5$, 2.0, 5.0, and $10$, 
where
steps occur at $R = 0.0$,
0.33, 0.5, and $0.66$.
For $\alpha_{d} = 5$ and $\alpha_d=10$, the
system remains locked in the $R = 0.0$ state over the
entire range of drives we have investigated.

\begin{figure}
  \begin{center}
    \includegraphics[width=0.6\columnwidth]{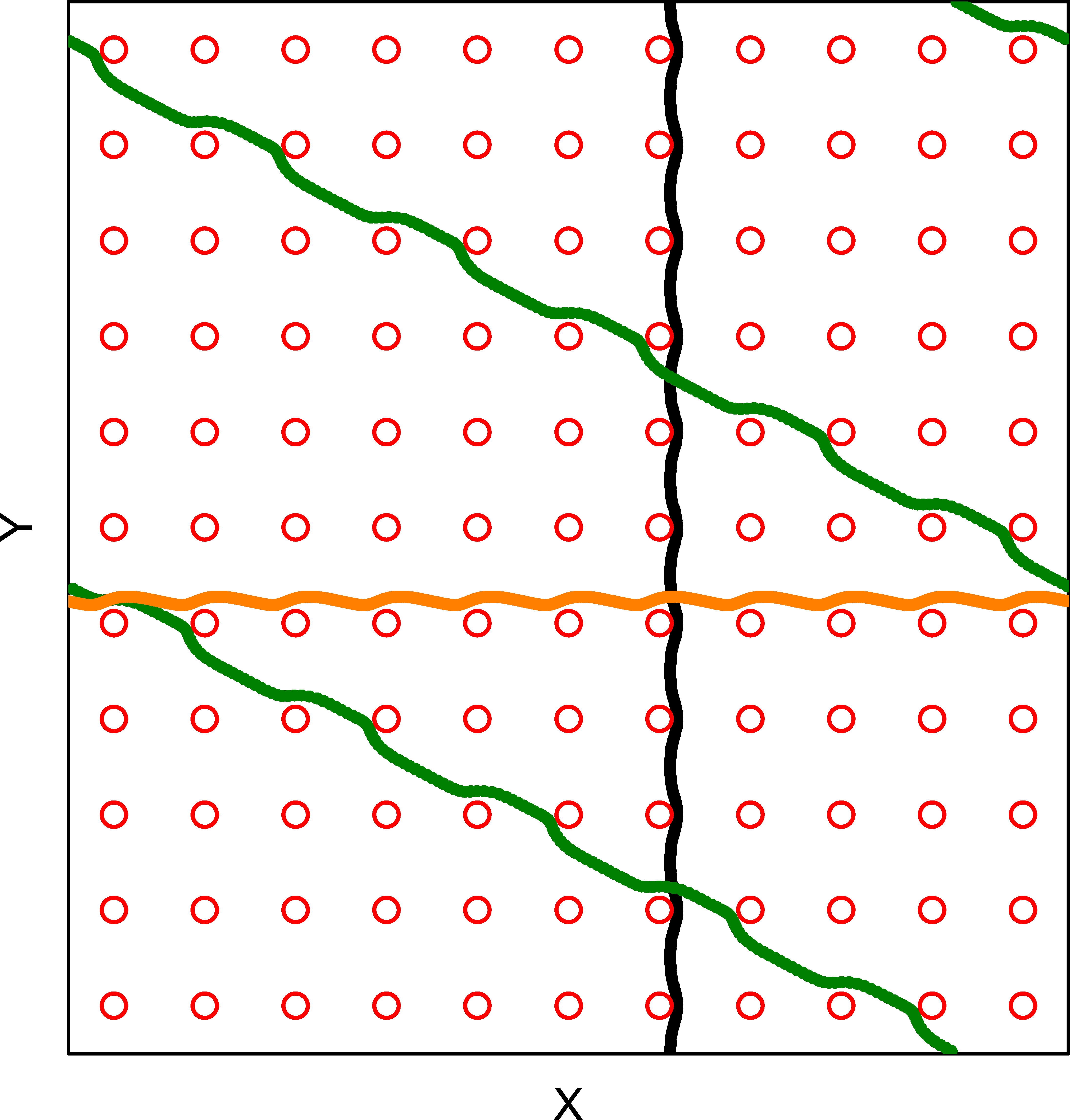}
    \end{center}
\caption{
  The obstacles (open circles) and the skyrmion trajectories (lines)
  for a system with $a_{0} = 0.65$ at $\alpha_{d} = 0.0$ (black), 2.0 (green),
  and $5.0$ (orange),
  where for
$\alpha_{d} = 0.0$ the skyrmion moves at $90^\circ$ with respect to the drive. 
}
\label{fig:12}
\end{figure}

In Fig.~\ref{fig:12} we plot the skyrmion trajectories for the
system in Figs.~\ref{fig:10} and \ref{fig:11} 
for
$\alpha_{d} = 0.0$,
2.0, and $5.0$ at $F^{D} = 1.0$.
In the case of zero damping with $\alpha_{d} = 0.0$, the 
skyrmion moves at $\theta_{sk} = 90^\circ$ with 
respect the drive.
For $\alpha_{d} = 5.0$ the skyrmion is locked 
to $\theta_{sk} = 0.0^\circ$,
while for $\alpha_{d} = 2.0$ the skyrmion moves at $R = -0.5$
corresponding to $\theta_{sk} = -26.56^\circ$,
indicating that increasing the damping reduces the magnitude of the
skyrmion Hall angle.

\begin{figure}
  \begin{center}
  \begin{minipage}{0.8\textwidth}
    \begin{minipage}{0.8\textwidth}
      \includegraphics[width=\textwidth]{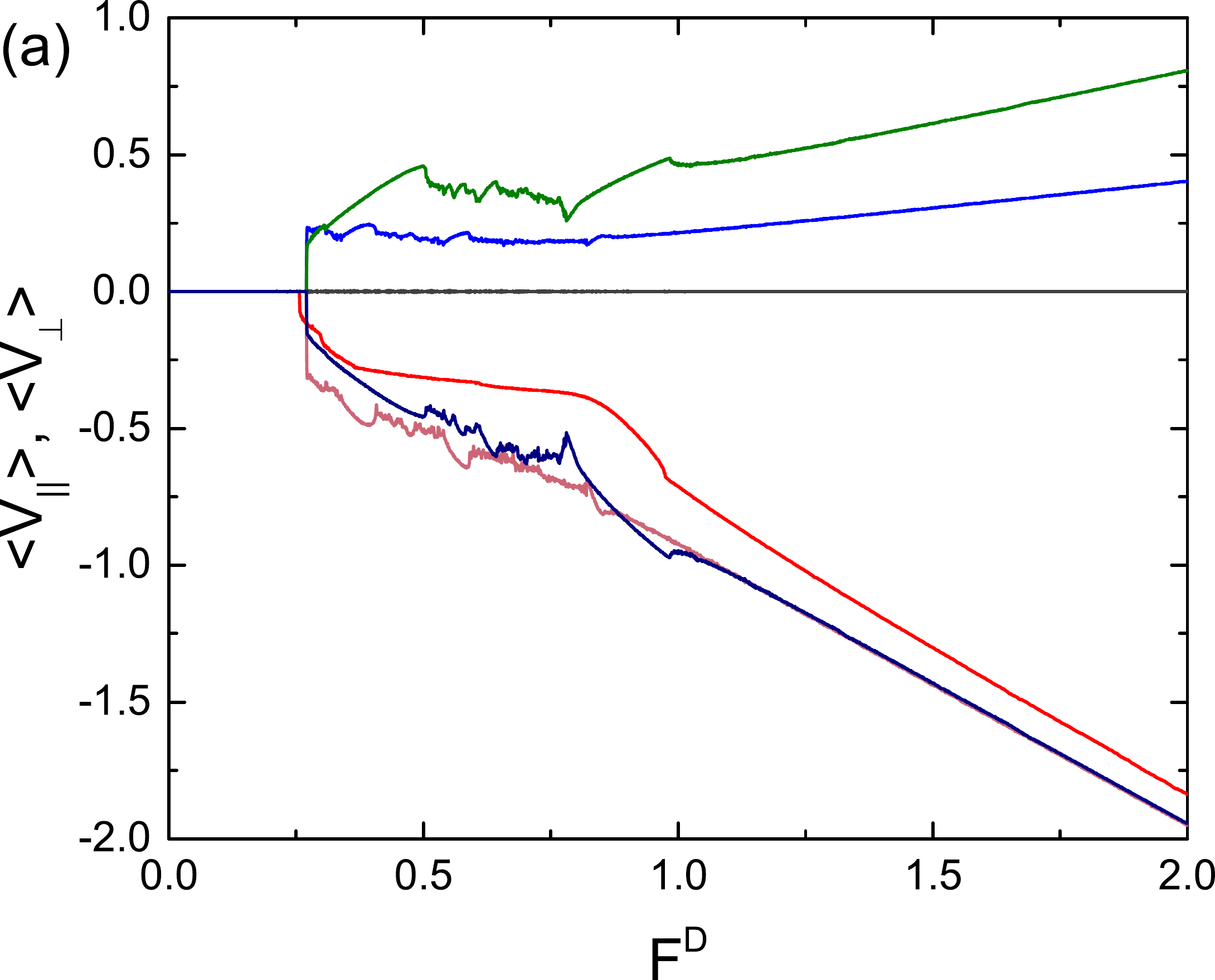}
    \end{minipage}\\
    \begin{minipage}{0.8\textwidth}
      \includegraphics[width=\textwidth]{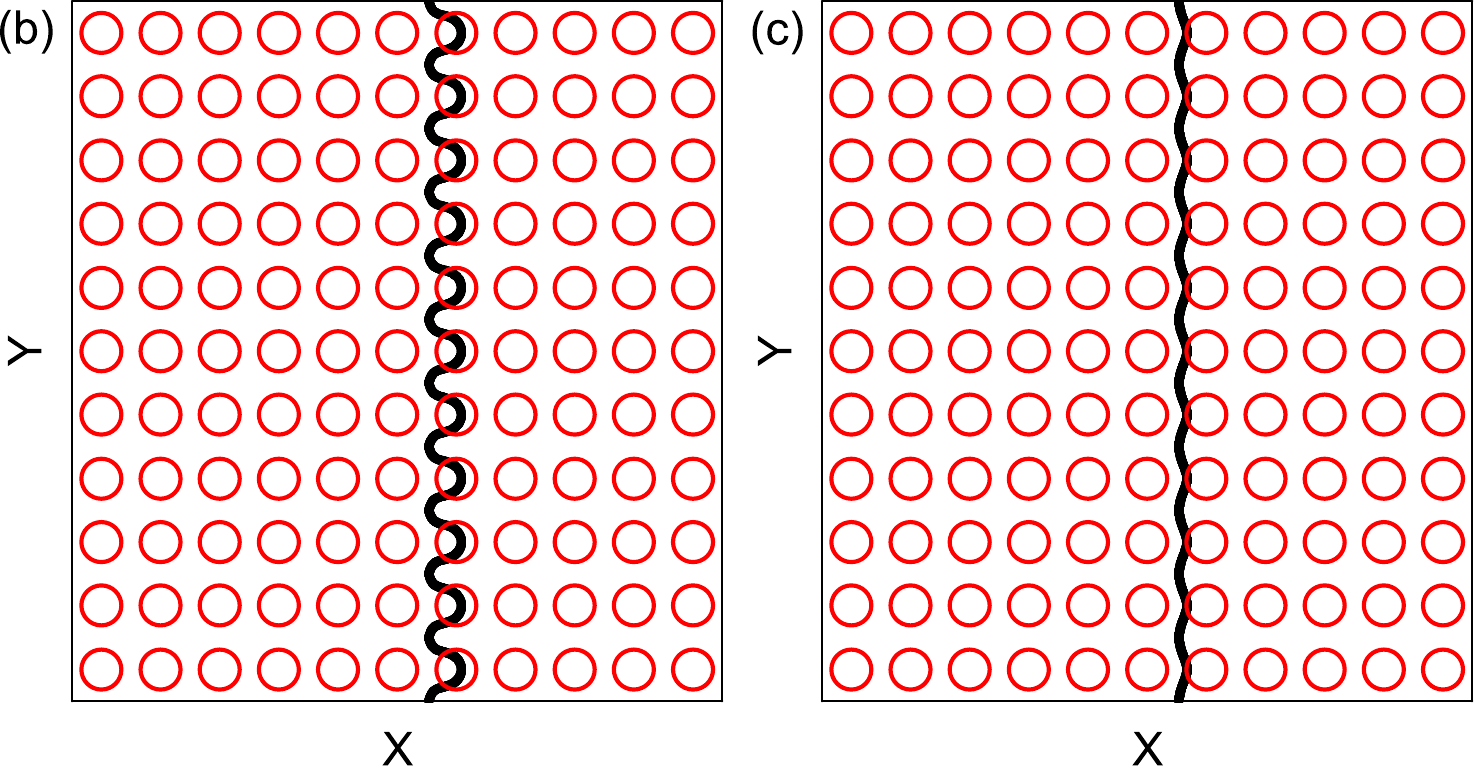}
    \end{minipage}
  \end{minipage}
  \end{center}
  \caption{
    (a) $\langle V_{||}\rangle$ (curves above zero) and $\langle V_{\perp}\rangle$ (curves
    below zero)
    vs $F^{D}$ for a system with
    $\alpha_{m} = 1.0$ at $\alpha_{d} = 0.0$
    (black, $\langle V_{||}\rangle$; red, $\langle V_{\perp}\rangle$);
    $\alpha_m=0.2$
    (light blue, $\langle V_{||}\rangle$; pink, $\langle V_{\perp}\rangle$);
    and $\alpha_m=0.4$
    (green, $\langle V_{||}\rangle$; dark blue, $\langle V_{\perp}\rangle$). 
    (b,c) The obstacles (open circles) and the skyrmion trajectory
    for the $\alpha_{d} = 0.0$ system at (b) $F^{D} = 0.5$ and (c) $F^D=1.0$.}
\label{fig:13}
\end{figure}

In Fig.~\ref{fig:13} we show $\langle V_{||}\rangle$ and $\langle V_{\perp}\rangle$
versus $F^D$
for a system 
with $a_{0} = 1.3$ at $\alpha_{d} = 0.0$, 0.2, and $0.4$. 
For  $\alpha_{d} = 0.0$, $\langle V_{||}\rangle = 0.0$ for all $F^{D}$. In this case
we observe a finite depinning threshold at $F^{D} = 0.255$,
where $\langle V_{\perp}\rangle$ first rises above zero.
There is also a partial two step depinning process,
with the second depinning transition occurring at
$F^{D} = 0.976$. 
In Fig.~\ref{fig:13}(b,c) we
illustrate the skyrmion trajectory in the system with $\alpha_{d} = 0.0$.
At $F^{D} = 0.5$, shown in Fig.~\ref{fig:13}(b),
the skyrmion is oscillating between the obstacles, while
in Fig.~\ref{fig:13}(c) at $F^D=1.0$,  above the second step,
the skyrmion runs along the edge of a column of obstacles.
At $\alpha_d=0.2$
in Fig.~\ref{fig:13}(a),
there is still a sharp depinning transition
along with 
a series of 
jumps in both $\langle V_{||}\rangle$ and $\langle V_{\perp}\rangle$.
We also find
an extended interval of $F^{D}$ over which
$\langle V_{||}\rangle$ decreases with increasing $F^D$,
but for $F^D>1.0$, $\langle V_{||}\rangle$ increases monotonically
with increasing $F^D$.
At $\alpha_{d} = 0.4$, there are a larger number of
jumps in the transport curves.

\begin{figure}
  \begin{center}
    \includegraphics[width=0.6\columnwidth]{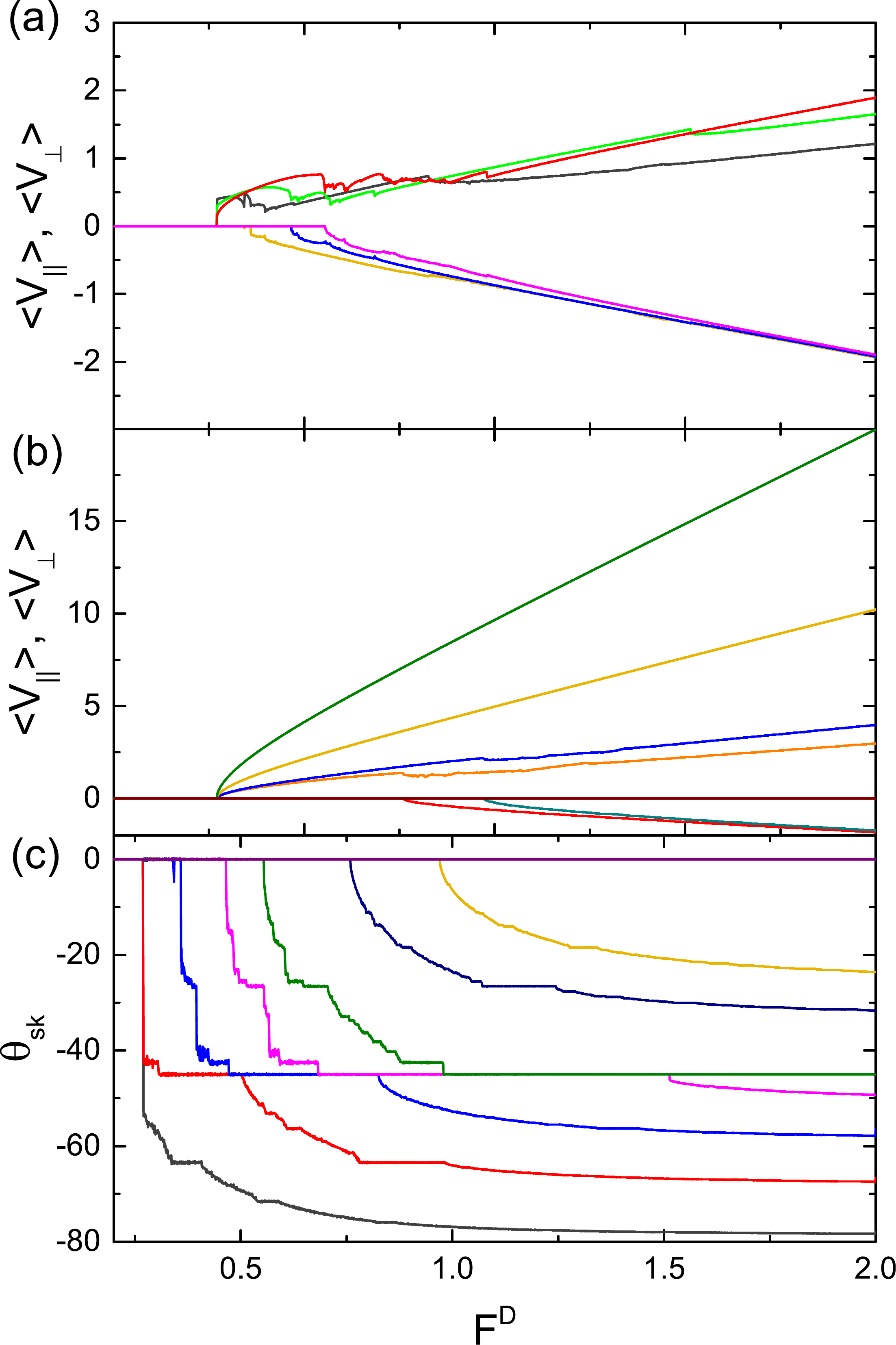}
    \end{center}
\caption{
  (a,b) $\langle V_{||}\rangle$ (curves above zero) and $\langle V_{\perp}\rangle$
  (curves below zero) vs $F^D$
  for a system with $\alpha_{m} = 1.0$ and $a_{0}= 1.3$.
  (a) $\alpha_{d} = 0.6$
  (black: $\langle V_{||}\rangle$; gold: $\langle V_{\perp}\rangle$);
  $\alpha_d=0.8$
  (light green: $\langle V_{||}\rangle$; blue: $\langle V_{\perp}\rangle$);
  and $\alpha_d=1.0$
  (red: $\langle V_{||}\rangle$; magenta: $\langle V_{\perp}\rangle$).
  (b) $\alpha_{d} = 1.5$
  (orange: $\langle V_{||}\rangle$; light red: $\langle V_{\perp}\rangle$);
  $\alpha_d=2.0$,
  (blue: $\langle V_{||}\rangle$; teal: $\langle V_{\perp}\rangle$);
  $\alpha_d=5.0$
  (gold: $\langle V_{||}\rangle$; brown: $\langle V_{\perp}\rangle$);
  and $\alpha_d=10$
  (green: $\langle V_{||}\rangle$; dark red: $\langle V_{\perp}\rangle$).
  (c) $\theta_{sk}$ vs $F^{D}$ for the samples in panels (a) and (b)
  with
  $\alpha_d=0.2$ (black),
  0.4 (red),
  0.6 (light blue),
  0.8 (magenta),
  1.0 (green),
  1.5 (dark blue),
  2.0 (gold),
  5.0 (light green),
  and 10.0 (purple).  Note that
  the curves for $\alpha_d=5.0$ and 10.0 are overlapping.
  }
\label{fig:14}
\end{figure}

In Fig.~\ref{fig:14}(a) we plot $\langle V_{||}\rangle$ and $\langle V_{\perp}\rangle$
versus $F^D$
for a system with $\alpha_m=1.0$ and $a_{0} = 1.3$
at $\alpha_{d} = 0.6$, 0.8, and $1.0$.
There is a finite
depinning threshold for motion in both the
parallel and perpendicular directions,
and the width of the $R = 0.0$ step grows 
with increasing damping.  
We show $\langle V_{||}\rangle$ and $\langle V_{\perp}\rangle$
versus $F^D$
for $\alpha_{d} = 1.5$, 2.0, 5.0, and $10$ in
Fig.~\ref{fig:14}(b),
where we find that the depinning transition is still sharp and that
the number of steps in the velocity-force curves decreases with
increasing damping.
In general, we find that for the drive at which
$\langle V_{\perp}\rangle$ becomes finite,
the magnitude of $\langle V_{||}\rangle$ drops.  
Figure~\ref{fig:14}(c) shows
$\theta_{sk}$ versus $F^{D}$ for the
samples in Figs.~\ref{fig:14}(a,b) where a series of locking
steps occur.  
In the case of $\alpha_{d} = 0.2$,
there are steps at $R = -2.0$, -3.0, and $-4.0$.  

Our results indicate that the skyrmion
Hall angle can be controlled precisely in systems with periodic substrates. 
This also implies that it may be possible to create new 
types of transistor-like devices
by exploiting the sharp jumps between the different Hall angles.
In this case, 
certain values of $R$ or $\theta_{sk}$
could be used to achieve
a specific resistance value.
It could also be possible to create
periodic arrays with obstacles of different sizes that would allow skyrmions to 
have one value of $\theta_{sk}$ in one part of the sample and a
different value of $\theta_{sk}$ in a different part of the sample.

\subsection{Attractive Pinning Sites}

\begin{figure}
  \begin{center}
    \includegraphics[width=0.6\columnwidth]{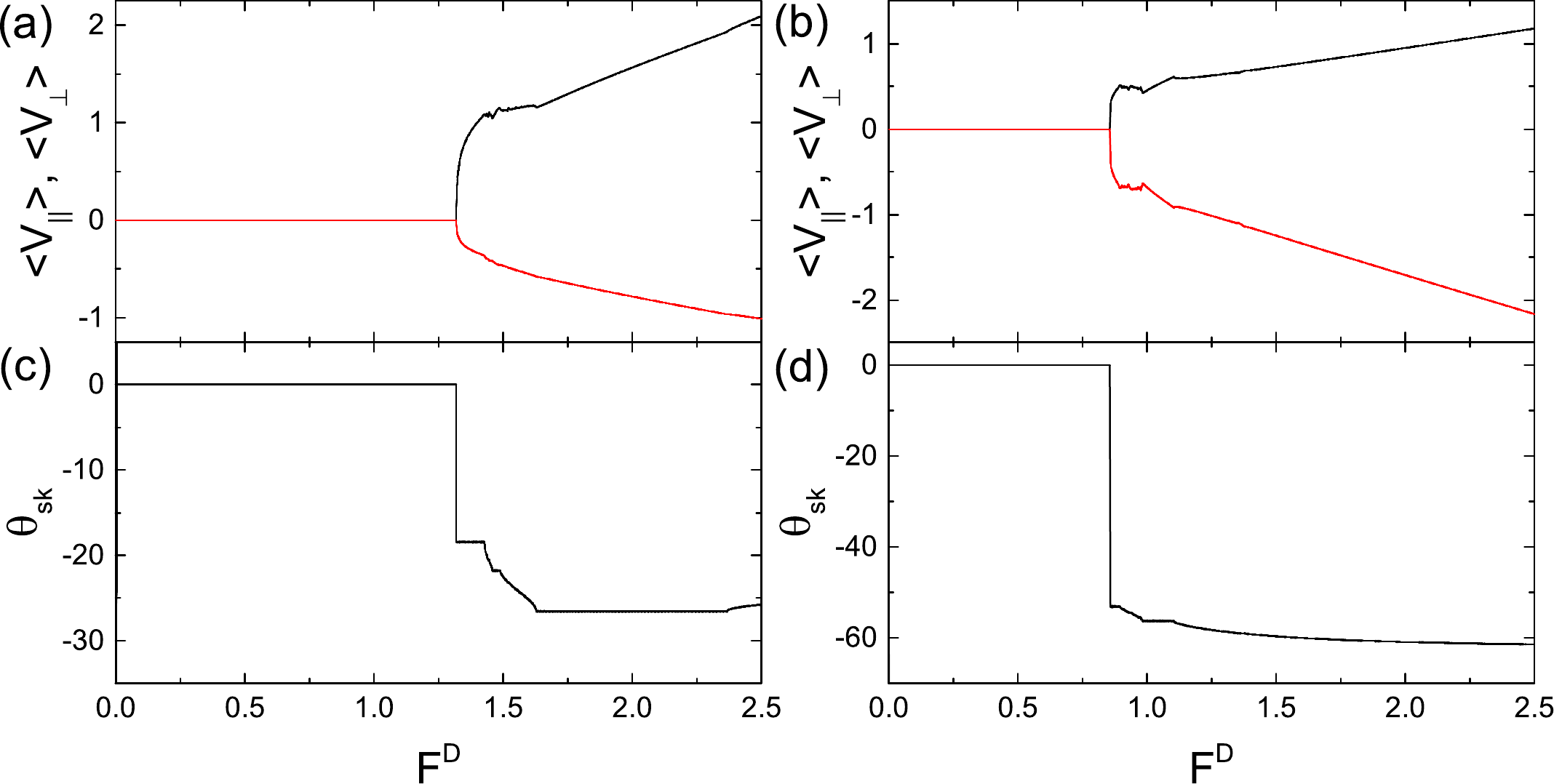}
    \end{center}
\caption{(a) $\langle V_{||}\rangle$ (black) and $\langle V_{\perp}\rangle$ (red)
  vs $F^{D}$ for the system in Fig.~\ref{fig:2}
  with $a_{0} = 0.65$ and $\alpha_{m}/\alpha_{d} = 0.45$
  but for attractive pinning sites.  
  When $F^{D} < 1.25$, the system is in a pinned state. 
  (b) $\langle V_{||}\rangle$ (black) and $\langle V_{\perp}\rangle$ (red)
  for an attractive pinning site system with
  $\alpha_{m}/\alpha_{d} = 1.91$ and $a_{0} = 1.0$.  
  (c) The skyrmion Hall angle $\theta_{sk}$  vs $F^{D}$ for the system in panel (a),
  where $R = \langle V_{\perp}\rangle /\langle V_{||}\rangle.$
(d) $\theta_{sk}$ vs $F^D$ for the system in panel (b).
}
\label{fig:New15}
\end{figure}

We next consider the effect of changing the substrate from obstacles to attractive
pinning sites,  which we achieve by changing the sign of $C_o$, the strength
of the potential.
A single skyrmion not subjected to a drive will come to rest inside such a
pinning site,
and there is a finite depinning threshold or driving force required
to set the skyrmion in motion. 
In Fig.~\ref{fig:New15}(a) we plot
$\langle V_{||}\rangle$
and $\langle V_{\perp}\rangle$
versus $F^{D}$ for the same system
as in Fig.~\ref{fig:2} containing a single skyrmion where
$a_{0} = 0.65$ and $\alpha_{m}/\alpha_{d} = 0.45$,
but for attractive pinning sites.
Figure~\ref{fig:New15}(c) shows the
corresponding skyrmion Hall angle.
Unlike the system with obstacles, where the skyrmions can initially slide along
the $x$ direction at low drives, for the attractive pins
we find a pinned regime
in which the skyrmion velocity is zero in both directions.
There are still
steps in the velocity-force curves and
in $\theta_{sk}$ but they are strongly reduced in size compared
to the system with obstacles.
In Fig.~\ref{fig:New15}(b) we plot $\langle V_{||}\rangle$ and $\langle V_{\perp}\rangle$
for the same system but at $\alpha_{m}/\alpha_{d} = 1.91$ and
$a_{0} = 1.0$.
Here the depinning threshold is lower, and although steps appear in the velocities
and in the skyrmion Hall angle shown in Fig.~\ref{fig:New15}(d),
they are strongly reduced in size.
This result indicates
that
the attractive pinning sites produce
much weaker directional locking effects than the obstacles. 

\section{Collective Effects and Topological Sorting}

\begin{figure}
  \begin{center}
    \includegraphics[width=0.6\columnwidth]{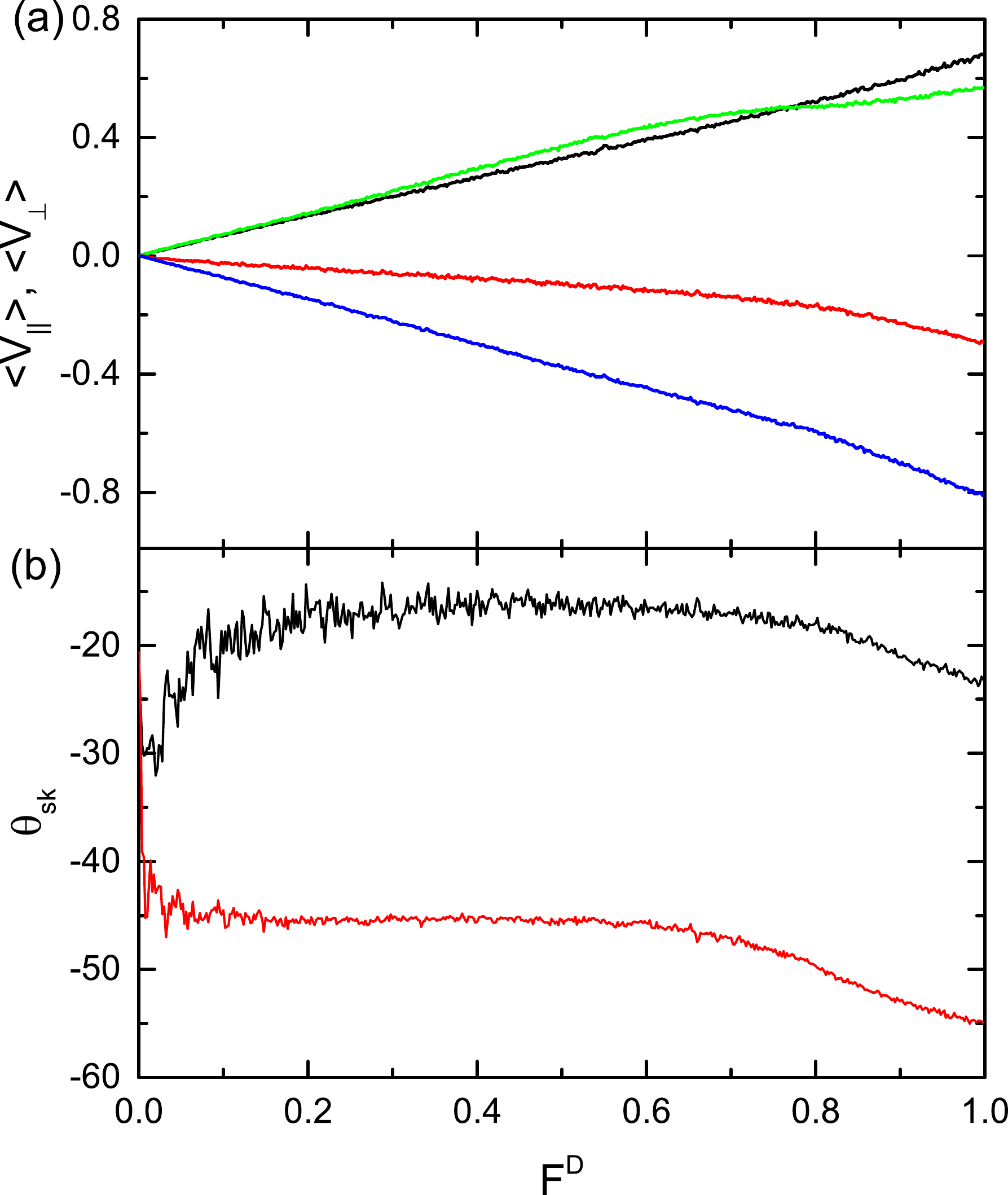}
  \end{center}
\caption{ (a) $\langle V_{||}\rangle$ (curves above zero) and $\langle V_{\perp}\rangle$
  (curves below zero) for each individual species
  versus $F^{D}$ for a system with two different species 
  of skyrmions at a total $n_s=0.034$.  Species $a$ has
  $\alpha_{m}/\alpha_{d} = 0.45$
  (black: $\langle V_{||}\rangle$; red: $\langle V_{\perp}\rangle$)
  and species $b$ has $\alpha_{m}/\alpha_{d} =  1.91$
  (green: $\langle V_{||}\rangle$; blue: $\langle V_{\perp}\rangle$).
  (b) The corresponding $R$ vs $F^{D}$ curves for the two species,
  $a$ (black) and $b$ (red).
}
\label{fig:15}
\end{figure}

\begin{figure}
  \begin{center}
    \includegraphics[width=0.5\columnwidth]{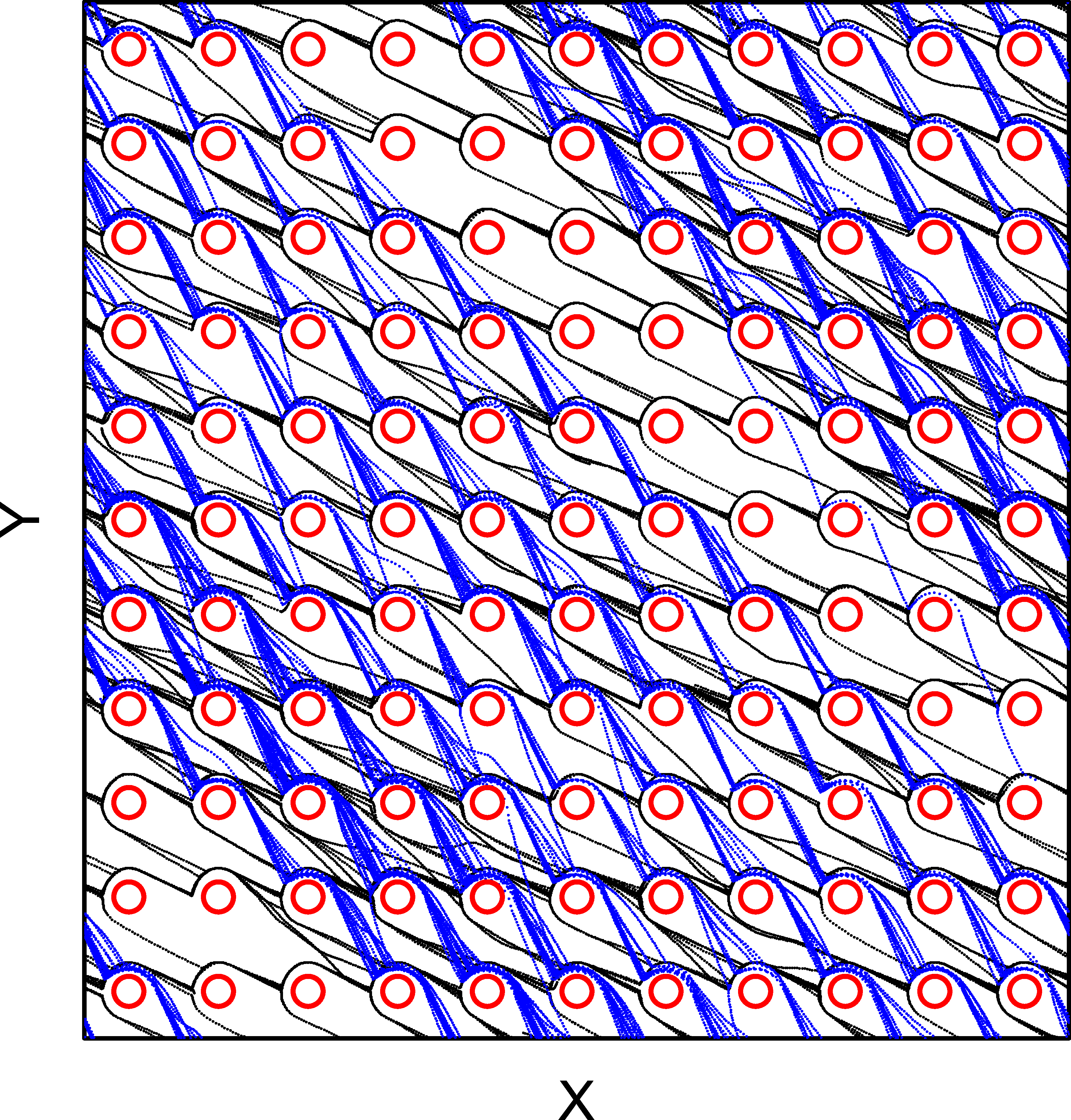}
    \end{center}
\caption{
  The obstacles (open circles) and skyrmion trajectories for
  species $a$ (black lines) and species $b$ (blue lines)
  for the system in Fig.~\ref{fig:15} at $F_{D} = 0.5$,
  where skyrmion species $a$ is locked to the $R = -1/3$ state and
  species $b$ is locked to the $R=-1$ state.
}
\label{fig:16}
\end{figure}

\subsection{Sorting}
We next consider
the effects of multiple interacting skyrmions moving in
periodic obstacle arrays for varied species of skyrmions.
In Fig.~\ref{fig:15}(a) we plot $\langle V_{||}\rangle$ and $\langle V_{\perp}\rangle$
for each species
versus $F^D$
for a system with $a_0=0.65$ containing 44 skyrmions
at a density of $n_{s} = 0.034$.
Half of the skyrmions 
have $\alpha_{m}/\alpha_{d} = 0.45$,
which we refer to as species $a$,
and the other half have
$\alpha_{m}/\alpha_{d} = 1.91$, which we refer to as species $b$.
Figure~\ref{fig:15}(b) shows the corresponding $R$
versus $F^D$ curves for the two species.
Here each species has a different value of $R$ at all drives.
We find that
$|R|$ is always lower than the
intrinsic value for either species, indicating that
the skyrmion-skyrmion
interactions cause a reduction in the Hall angle for both species.
The pronounced steps associated with locking effects
are not present in this system.
This is due to the disordering effect that occurs
when the different skyrmion species each move at different
angles, causing collisions among the skyrmions and resulting in the
disordered trajectories illustrated in
Fig.~\ref{fig:16} 
for $F^{D} = 0.5$.
Here,
species $a$
is moving at $R = -1/3$,
but when the species $a$ skyrmions collide
with the species $b$ skyrmions, the trajectories begin to disorder.
As $F^{D}$ is increased, species $a$ eventually jumps out of the
$R = -1/3$ state and moves at a higher angle.
For collective effects among monodisperse skyrmions,
locking steps similar to
those found
for the single skyrmion case occur.
The fact that different skyrmion species can lock to different
directions of motion indicates that
it would be possible to perform
skyrmion sorting
in systems containing skyrmions with different sizes or different winding numbers,
similar to the species fractionation that can be achieved for
different species of colloidal particles
moving over periodic substrates
\cite{MacDonald03,Lacasta05,Herrmann09,Ladavac04}. 
The separation will be more difficult in cases where there are
only small differences between the skyrmion species,
but it should be possible to tune $F^D$ carefully
to the edge of a locking step such that one species is in a locked state
while the other species is not.

\subsection{Jamming Effect}

\begin{figure}
  \begin{center}
    \includegraphics[width=0.6\columnwidth]{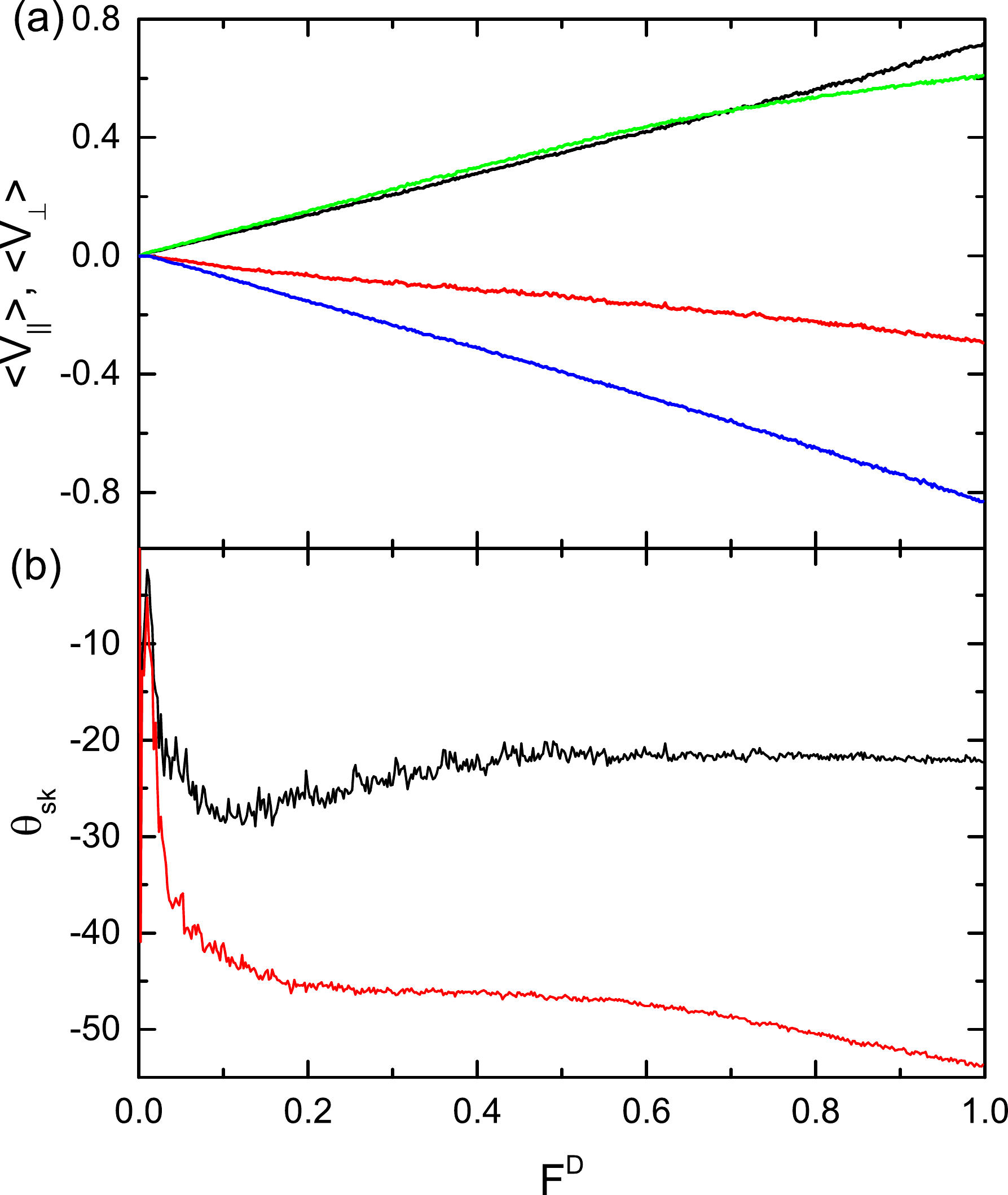}
  \end{center}
\caption{
  (a) $\langle V_{||}\rangle$ (curves above zero) and $\langle V_{\perp}\rangle$
  (curves below zero)
  for species $a$
  (black: $\langle V_{||}\rangle$; red: $\langle V_{\perp}\rangle$)
  and $b$
  (green: $\langle V_{||}\rangle$; blue: $\langle V_{\perp}\rangle$)
  in a system with $n_{s} = 0.085$.
  (b) The corresponding $R$ vs $F^{D}$ for
  species $a$ (black) and $b$ (red).
}
\label{fig:17}
\end{figure}

\begin{figure}
  \begin{center}
  \begin{minipage}{0.6\columnwidth}
    \begin{minipage}{0.6\columnwidth}
      \includegraphics[width=\textwidth]{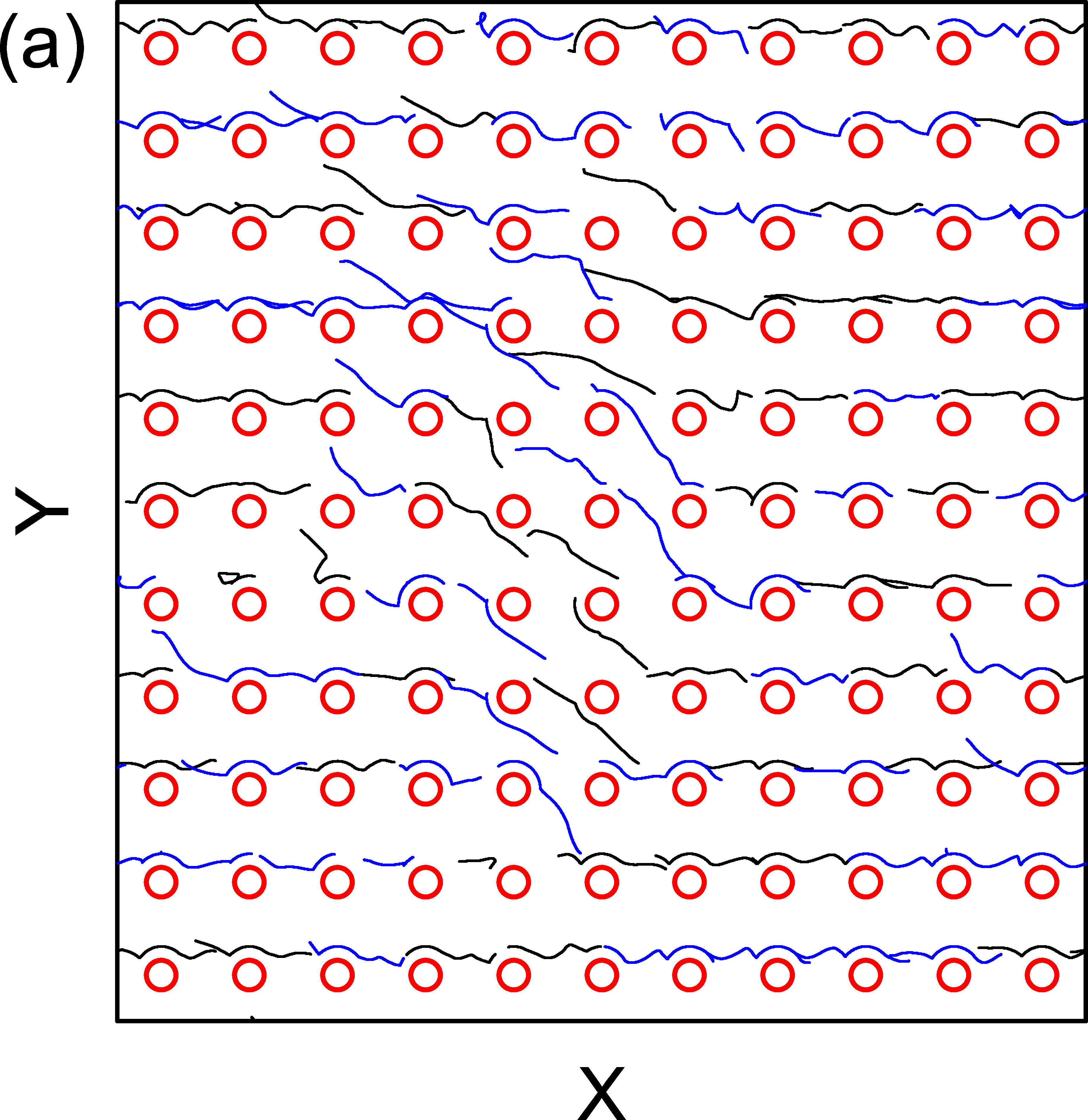}
    \end{minipage}%
    \begin{minipage}{0.6\columnwidth}
      \includegraphics[width=\textwidth]{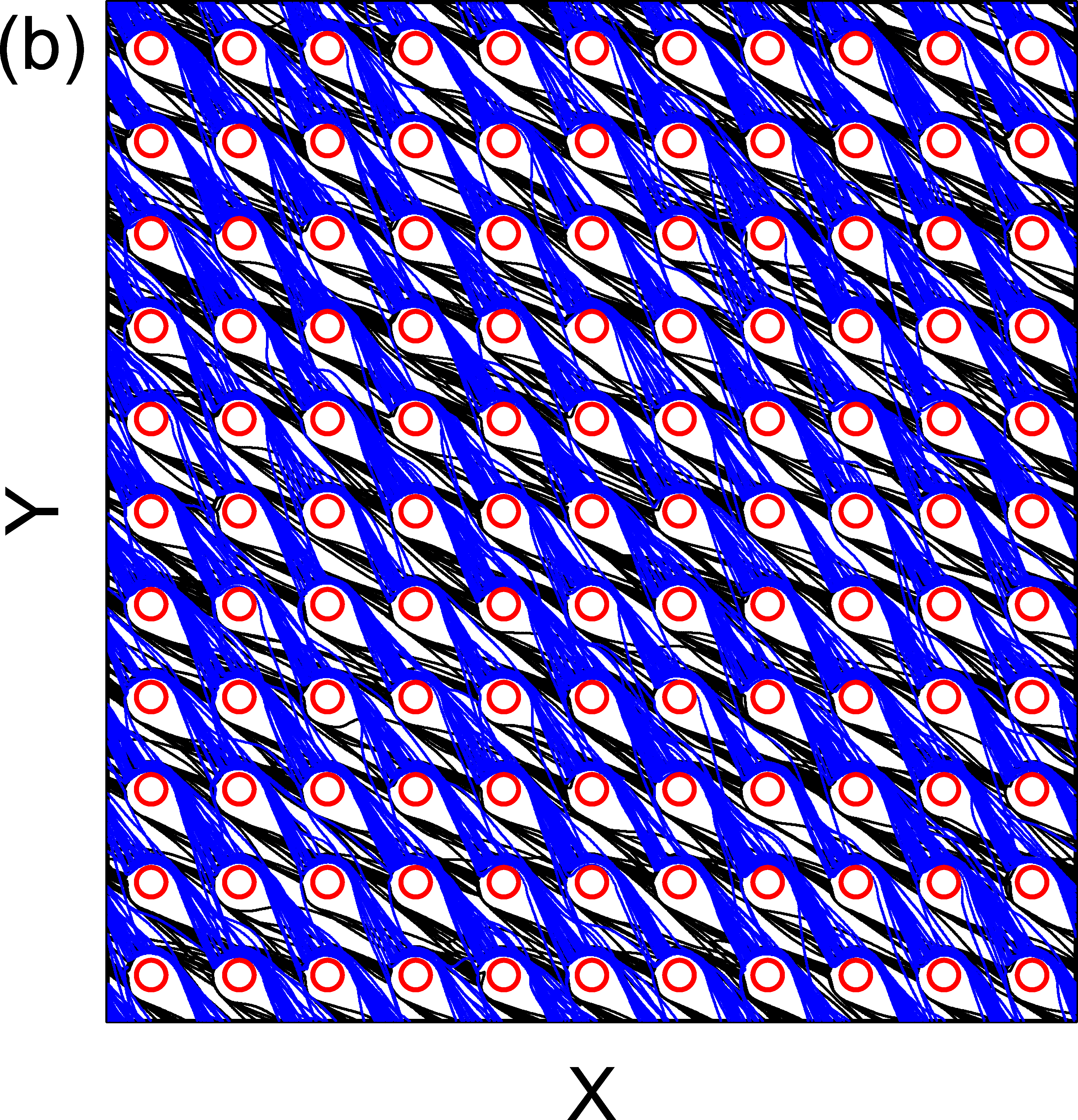}
    \end{minipage}
  \end{minipage}
  \end{center}
  \caption{
    The obstacles (open circles) and skyrmion trajectories for species $a$ (black lines)
    and species $b$ (blue lines) for the system in Fig.~\ref{fig:17}.
    (a) At $F^{D} = 0.01$, the system forms a jammed phase
close to $R = 0.0$ where all of the skyrmions move together in elastic flow.
(b) At $F^{D} = 0.5$, species $b$ is locked to the $R = -1.0$ state and
species $a$ is partially locked to the $R  = -0.33$ state.  
}
\label{fig:18}
\end{figure}

As the density of the two species system increases,
the skyrmion-skyrmion interactions become stronger,
and low drive regimes emerge in which
the two species move while rigidly locked together.
In Fig.~\ref{fig:17}(a) we show $\langle V_{||}\rangle$ and $\langle V_{\perp}\rangle$
for each species versus $F^D$ in a sample with $n_s=0.085$,
and in Fig.~\ref{fig:17}(b) we plot the corresponding $R$ versus $F^{D}$ curves.
Species $a$ reaches a step with
$R = -0.5$, which is lower than 
the expected intrinsic value of $R = -0.45$,
indicating that there is dragging effect from species $b$
on species $a$.
At higher drives,
species $a$ settles onto a step with
$|R| = 0.4$.
At the lowest values of $F^{D}$, $R$ is close to zero
when the system
forms a jammed state where the skyrmions of both species
start to form an elastically moving lattice due to the repulsive
skyrmion-skyrmion interactions.
In Fig.~\ref{fig:18}(a) we show the skyrmion trajectories for the system in
Fig.~\ref{fig:17} at $F^{D} = 0.01$, where the skyrmions
move  mostly in the $R = 0.0$ state with some small jumps in the $y$-direction,
and where the two species remain 
locked together.
Figure~\ref{fig:18}(b) shows that at $F^{D} = 0.5$,
species $b$ is locked to
$R = -1.0$ and species $a$ is partially locked to $R = -1/3$.

\begin{figure}
  \begin{center}
    \includegraphics[width=0.6\columnwidth]{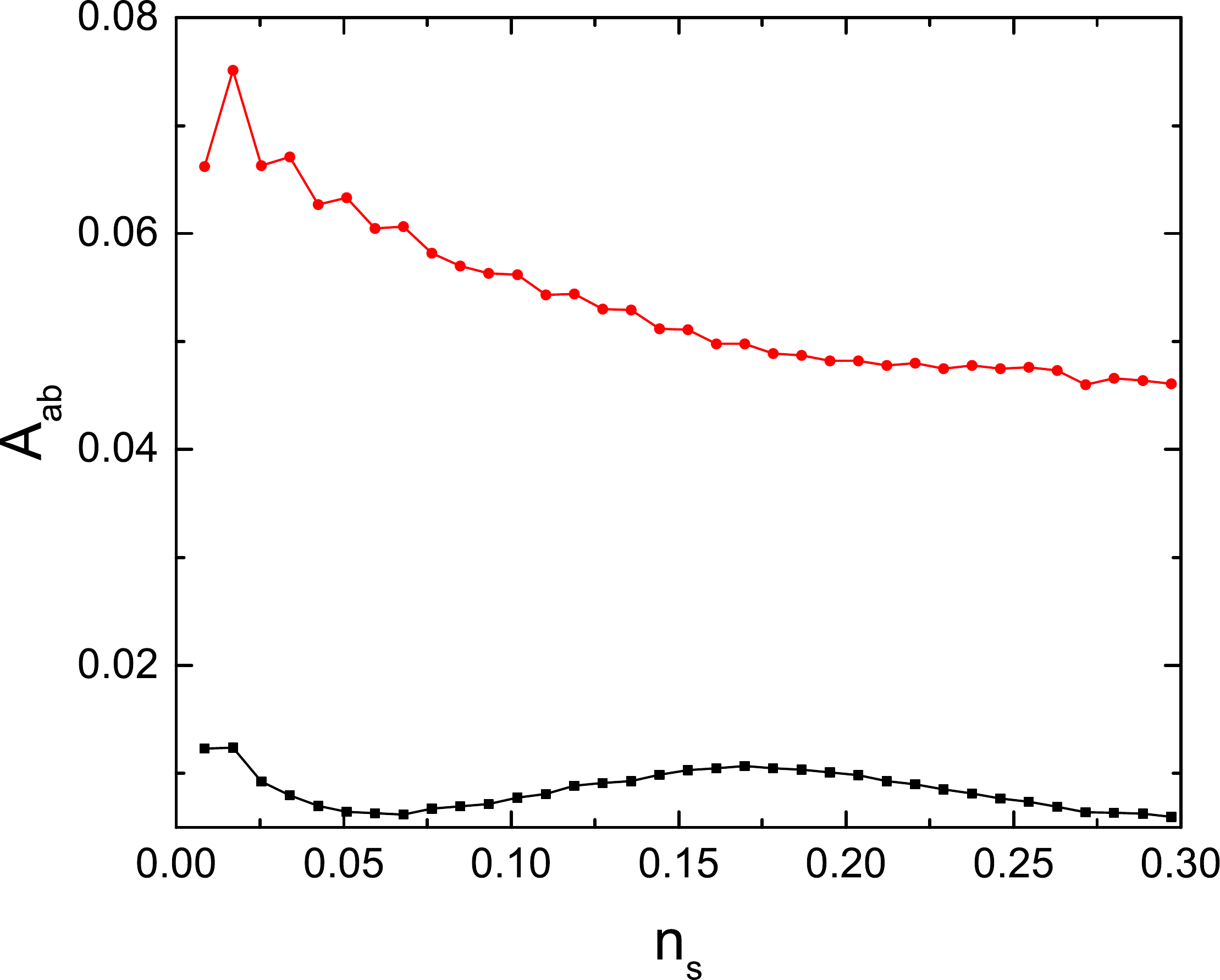}
    \end{center}
\caption{ 
  The area $A_{ab}$ between the velocity curves
  vs skyrmion density $n_{s}$ for
  $\langle V_{||}\rangle$ (black curve) and $\langle V_{\perp}\rangle$ (red curve),
  as a measure of the efficiency of the
  sorting of the two species. 
  The efficiency drops monotonically with increasing $n_s$ in the perpendicular
  direction but is non-monotonic for the
parallel direction. 
}
\label{fig:19}
\end{figure}

\begin{figure}
  \begin{center}
    \includegraphics[width=0.6\columnwidth]{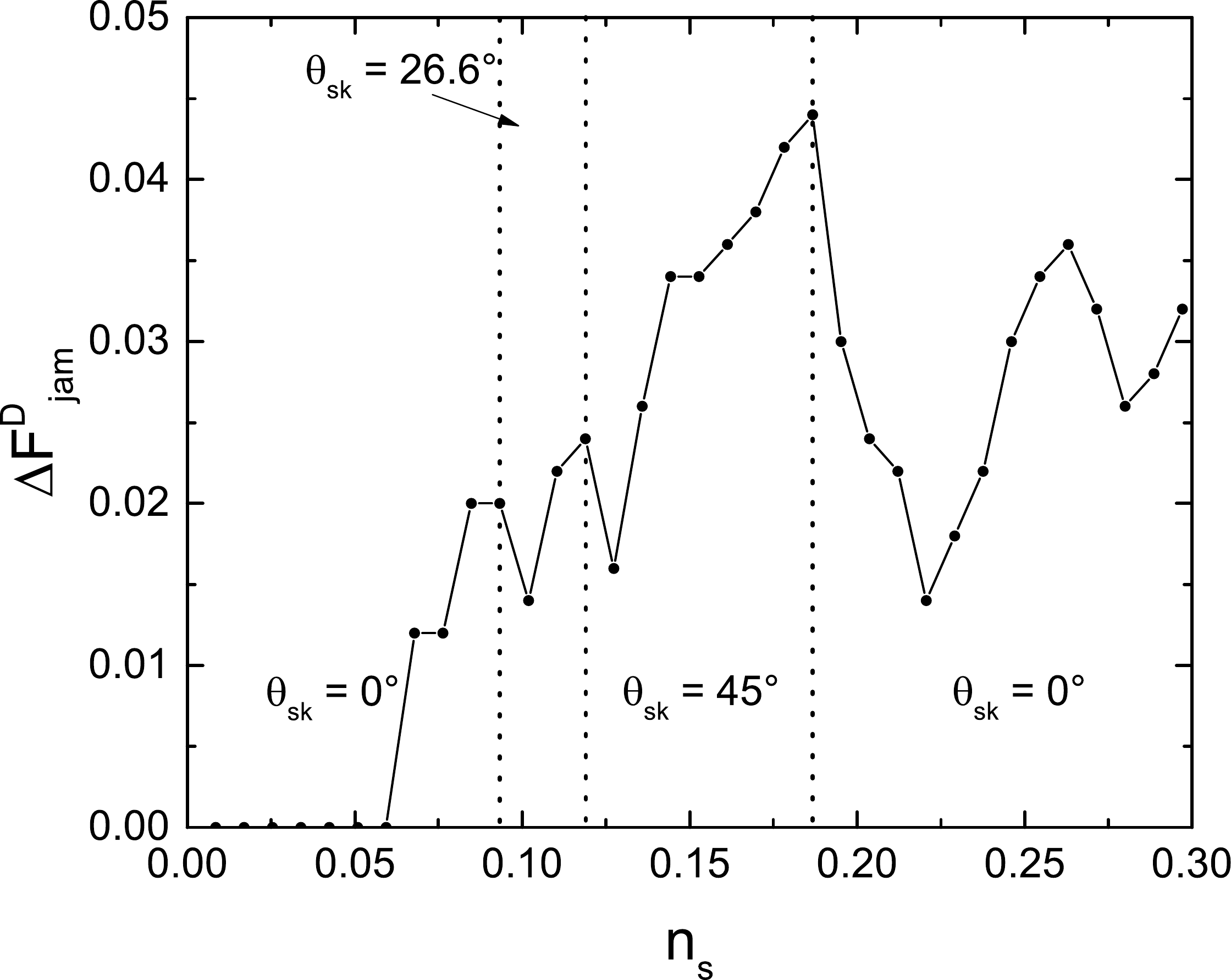}
    \end{center}
\caption{ 
  $\Delta F^{D}_{\rm jam}$, the width of the window in $F^D$ where jamming
  behavior occurs, vs skyrmion density $n_{s}$ at the lower drives,
  highlighting the efficiency 
  of the jammed phases.
  The vertical lines indicate the direction in which the jammed phase is moving
  for the $R = 0$, -0.5, and $-1.0$ states. 
}
\label{fig:20}
\end{figure}

To quantify the efficiency of the separation of the two species, we measure
the difference $A_{ab}=\langle V_a\rangle -\langle V_b\rangle$ between the
velocity of the two species in both the 
parallel and perpendicular directions.
In Fig.~\ref{fig:19} we show that
the efficiency $A_{ab}$ of the separation drops with increasing
skyrmion density for the
system from Figs.~\ref{fig:16} and {\ref{fig:17}.
For the perpendicular direction, there is a monotonic decrease in the
efficiency with increasing $n_s$
due to the increasing drag effect between the two species
that appears as the density increases, while for the parallel direction, 
the behavior of the efficiency is non-monotonic
as a function of $n_s$ due to a partial jamming effect.
The jamming effect appears at
low drives and can also occur for motion in different directions.
In Fig.~\ref{fig:20} 
we plot the width $\Delta F^D_{\rm jam}$ of the jammed phase as a function
of $n_s$, where the jammed phase is defined to extend from the depinning
threshold to the drive at which the two skyrmion species first begin to move at
different velocities.
For $n_{s} \leq 0.0594$, there is no jamming since the
skyrmion density is low enough that skyrmions
can easy pass by each other with minimal interactions.
For $ 0.0594 < n_{s} < 0.0933$, the initial motion occurs
in the $R = 0.0$ jammed
state, while
for $0.0933 < n_{s} < 0.1188$,
the skyrmions form a jammed state that moves in the $R = -0.5$ direction.
Over the interval
$0.1188<n_s<0.1867$,
the jammed state moves in the
$R = -1.0$ direction, while for 
$n_{s} > 0.1867$,
another jammed state appears that is locked in the $R = 0.0$ direction.
For larger values of $n_{s}$, the 
jammed states increase in extent but can move along different directions.

\section{Summary}
We have examined individual and multiple interacting
skyrmions moving through a square array of obstacles. 
In the single skyrmion case,
we observe a series of directional locking effects where the skyrmion Hall angle increases
in both a quantized and a continuous manner.
The transitions between the different locking steps are associated with
dips or cusps in the velocity-force curves as well as with a quantized skyrmion Hall angle. 
For small obstacles, the skyrmion motion is oriented
close to the intrinsic Hall angle, but directional locking to
higher or lower Hall angles can occur.
For larger obstacles, the number of directional locking steps is
increased.
The angle of skyrmion motion is 90$^\circ$ with respect to the drive
at zero damping, and it decreases with increasing damping until,
for high damping, the skyrmion 
remains locked in the drive direction.
When multiple species of interacting skyrmions are present,
we show that it is possible to achieve a
sorting effect in which one species of skyrmion locks to a symmetry direction
of the obstacle lattice while the other species does not.
The sharp steps in the velocity force curves disappear when there
are multiple skyrmion species due to the disordering of the skyrmion
trajectories that occurs when the different species move in different
directions and collide.
At lower drives, as the skyrmion density increases
we observe a jammed state
in which the two
species
form a rigid
assembly and all move in the same direction,
while at higher drives the motion of the two species decouples.
For increasing skyrmion density, we observe
a series of transitions among jammed phases
that move at different angles with respect to the drive.  

\ack
This work was supported by the US Department of Energy through
the Los Alamos National Laboratory.  Los Alamos National Laboratory is
operated by Triad National Security, LLC, for the National Nuclear Security
Administration of the U. S. Department of Energy (Contract No. 892333218NCA000001).
N.P.V. acknoledges
funding from
Funda\c{c}\~{a}o de Amparo \`{a} Pesquisa do Estado de S\~{a}o Paulo - FAPESP (Grant 2018/13198-7).

\section*{References}
\bibliographystyle{iopart-num}
\bibliography{mybib}
\end{document}